\documentclass[fleqn,usenatbib]{config/mnras}

\usepackage[T1]{fontenc}
\usepackage{lastpage}
\usepackage{ulem}
\usepackage{CJKutf8}

\DeclareRobustCommand{\VAN}[3]{#2}
\let\VANthebibliography\thebibliography
\def\thebibliography{\DeclareRobustCommand{\VAN}[3]{##3}\VANthebibliography}

\usepackage{amsmath}	
\usepackage{graphicx}
\usepackage{booktabs}
\usepackage{multirow}	
\usepackage{paralist}
\usepackage{textgreek}
\usepackage{xargs}
\usepackage[dvipsnames]{xcolor}
\usepackage{xspace}
\usepackage{txfonts}
\usepackage{hyperref}
\hypersetup{
    colorlinks=true,
    linkcolor=Cerulean,
    citecolor=NavyBlue,
    filecolor=magenta,
    urlcolor=Violet,
    pdftitle={The Gustav Mahler Galaxy and Alma Francesco D'Eugenio JADES JWST},
    }
\usepackage{subcaption}
\usepackage{adjustbox}  

\newcommand{\rekt}[1]{\text{\raisebox{0.37em}{\fbox{}}}\ensuremath{_{#1}}\xspace}

\graphicspath{{./}{figs/}}
\usepackage{etoolbox}
\makeatletter
\newcommand\sendemail[4]{
\edef\@tempa{mailto:#1?subject=#2&body=#3 }%
\edef\@tempb{\expandafter\html@spaces\@tempa\@empty}%
\href{\@tempb}{#4}}

\catcode\%=11
\def\html@spaces#1 #2{#1
\catcode\%=14
\makeatother

\newcommand{\todo}[1]{\textcolor{magenta}{[#1]}}
\newcommand{\orcid}[2]{\href{http://orcid.org/#2}{#1}}
\newcommand{\orcidsymb}[2]{\href{http://orcid.org/#2}{#1\adjustbox{trim={-.15\width} {0\height} {-.15\width} {0\height},clip}{\includegraphics[height=10pt]{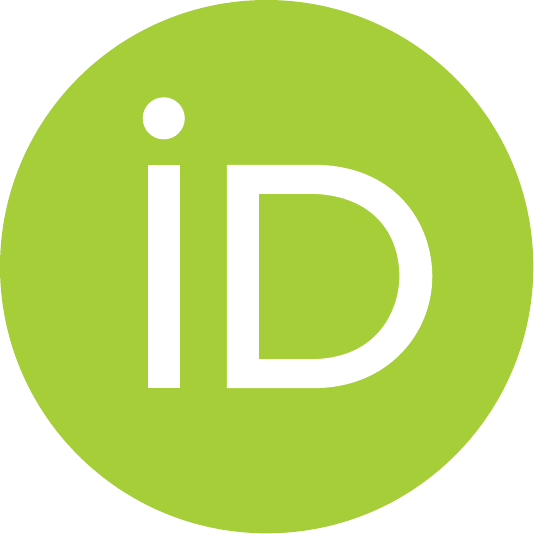}}}}

\newcommand{\citationneeded}{\textcolor{ForestGreen}{$^{\rm citation\;needed}$}}
\let\oldtextsigma\textsigma
\renewcommand{\textsigma}{\oldtextsigma\xspace}
\let\oldAA\AA
\renewcommand{\AA}{\text{\oldAA}\xspace}
\let\oldtextdegree\textdegree
\renewcommand{\textdegree}{\oldtextdegree\xspace}

\newcommand{\kms}{\ensuremath{\mathrm{km\,s^{-1}}}\xspace}
\newcommand{\Msun}{\ensuremath{{\rm M}_\odot}\xspace}
\newcommand{\Zsun}{\ensuremath{{\rm Z}_\odot}\xspace}
\newcommand{\yr}{\ensuremath{{\rm yr}}\xspace}
\newcommand{\Myr}{\ensuremath{{\rm Myr}}\xspace}
\newcommand{\Gyr}{\ensuremath{{\rm Gyr}}\xspace}
\newcommand{\peryr}{\ensuremath{{\rm yr^{-1}}}\xspace}
\newcommand{\Lsun}{\hbox{\,${\rm L}_\odot$}}
\newcommand{\mum}{\text{\textmu m}\xspace}
\newcommand{\kpc}{\text{kpc}\xspace}
\newcommand{\ZH}{\text{[Z/H]}\xspace}

\newcommandx{\lambdar}[2][1=R,2=]{\ensuremath{\lambda_{\rm {#1}}{#2}}\xspace}
\newcommand{\eps}{\ensuremath{\epsilon}\xspace}
\newcommand{\mstar}{\ensuremath{M_\star}\xspace}
\newcommand{\mdyn}{\ensuremath{M_\mathrm{dyn}}\xspace}
\newcommand{\re}{\ensuremath{R_\mathrm{e}}\xspace}
\newcommand{\vstar}{\ensuremath{v_\star}\xspace}
\newcommand{\vnai}{\ensuremath{v_{\NaI}}\xspace}
\newcommand{\sigmastar}{\ensuremath{\sigma_\star}\xspace}
\newcommand{\sigmaestar}{\ensuremath{\sigma_{\star,\mathrm{e}}}\xspace}
\newcommand{\vperc}[1]{\ensuremath{v_{#1}}\xspace}

\newcommand{\vesc}{\ensuremath{v_\mathrm{esc}}\xspace}
\newcommand{\nelec}{\ensuremath{n_\mathrm{e}}\xspace}
\newcommand{\Rout}{\ensuremath{R_\mathrm{out}}\xspace}
\newcommand{\vout}{\ensuremath{v_\mathrm{out}}\xspace}
\newcommandx{\Mout}[2][1=,2=]{\ensuremath{M_{\mathrm{out}{#2}}^{#1}}\xspace}
\newcommandx{\Mdotout}[2][1=,2=]{\ensuremath{\dot{M}_{\mathrm{out}{#2}}^{#1}}\xspace}

\newcommandx{\fluxdcgs}[1][1=-20]{$\times 10^{[#1]}$~erg~s$^{-1}$~cm$^{-2}$~\AA$^{-1}$\xspace}
\newcommandx{\fluxcgs}[2][1=-20,2=\ensuremath{\times}]{${#2}10^{#1}$~erg~s$^{-1}$~cm$^{-2}$\xspace}
\newcommandx{\powercgs}[1][1=44]{$\times 10^{#1}$~erg~s$^{-1}$\xspace}
\newcommand{\Av}{\ensuremath{A_V}\xspace}
\newcommandx{\sbcgs}[1][1=-18]{$\times 10^{#1}$~erg~s$^{-1}$~cm$^{-2}$~arcsec$^{-2}$\xspace}

\newcommand{\fHbetash}{$f(\mathrm{H}\text{\textbeta})_\mathrm{sh}$\xspace}
\newcommand{\target}{\text{JADES-GS-518794}\xspace}
\newcommand{\satellite}{\text{JADES-GS-518709}\xspace}

\newcommand{\jwst}{\textit{JWST}\xspace}
\newcommand{\hst}{\textit{HST}\xspace}
\newcommand{\ppxf}{{\sc ppxf}\xspace}
\newcommand{\prospector}{{\sc prospector}\xspace}
\newcommand{\prospectorbeta}{{\sc prospector}\,\textbeta\xspace}
\newcommand{\emcee}{{\sc emcee}\xspace}
\newcommand{\cloudy}{{\sc cloudy}\xspace}
\newcommand{\pyneb}{{\sc pyneb}\xspace}
\newcommand{\pysersic}{{\sc pysersic}\xspace}
\newcommandx{\mappings}[1][1=]{{\sc mappings{#1}}\xspace}
\newcommand{\eazy}{{\sc eazy}\xspace}
\newcommand{\cigale}{{\sc cigale}\xspace}

\newcommand{\Mdynvalue}{$\Mdyn = 2.0\pm0.5 \times 10^{11}$~\MSun}

\defcitealias{tacchella+2022a}{T22}
\defcitealias{nesvadba+2017}{N17}

\newcommand{\Lyalpha}{\text{Ly\,\textalpha}\xspace}
\newcommand{\Halpha}{\text{H\,\textalpha}\xspace}
\newcommand{\Hbeta}{\text{H\,\textbeta}\xspace}
\newcommand{\Hgamma}{\text{H\,\textgamma}\xspace}
\newcommand{\Hdelta}{\text{H\,\textdelta}\xspace}
\newcommand{\Paalpha}{\text{Pa\,\textalpha}\xspace}
\newcommand{\Pabeta}{\text{Pa\,\textbeta}\xspace}
\newcommand{\Hepsilon}{\text{H\,\textepsilon}\xspace}

\newcommandx{\permittedEL}[6][1=O,2=III,3=,4=,5=,6=]{\text{{#1}\,{\sc {#2}}{#3}{#4}{#5}{#6}}\xspace}
\newcommandx{\semiforbiddenEL}[6][1=O,2=III,3=,4=,5=,6=]{\text{{#1}\,{\sc{#2}}]{#3}{#4}{#5}{#6}}\xspace}
\newcommandx{\forbiddenEL}[6][1=O,2=III,3=,4=,5=,6=]{\text{[{#1}\,{\sc{#2}}]{#3}{#4}{#5}{#6}}\xspace}

\newcommand{\EW}[1]{\text{EW(#1)}\xspace}

\newcommand{\HI}{\permittedEL[H][i]}
\newcommand{\HII}{\permittedEL[H][ii]}

\newcommand{\NV}{\permittedEL[N][v]}
\newcommandx{\NVL}[1][1=1243]{\permittedEL[N][v][\textlambda][#1]}
\newcommandx{\NVall}{\permittedEL[N][v][\textlambda][\textlambda][1239,][1243]}

\newcommandx{\CIIL}[1][1=232x]{\semiforbiddenEL[C][ii][\textlambda][#1]}
\newcommandx{\CIIall}{\semiforbiddenEL[C][ii][\textlambda][\textlambda][2324--][2329]}

\newcommand{\NIV}{\semiforbiddenEL[N][iv]}
\newcommandx{\NIVL}[1][1=1486]{\semiforbiddenEL[N][iv][\textlambda][#1]}

\newcommand{\CIV}{\permittedEL[C][iv]}
\newcommandx{\CIVL}[1][1=1550]{\permittedEL[C][iv][\textlambda][#1]}
\newcommand{\CIVall}{\permittedEL[C][iv][\textlambda][\textlambda][1548,][1551]}

\newcommand{\HeII}{\permittedEL[He][ii]}
\newcommandx{\HeIIL}[1][1=1640]{\permittedEL[He][ii][\textlambda][#1]}

\newcommand{\semiOIII}{\semiforbiddenEL[O][iii]}
\newcommandx{\semiOIIIL}[1][1=1666]{\semiforbiddenEL[O][iii][\textlambda][#1]}
\newcommand{\semiOIIIall}{\semiforbiddenEL[O][iii][\textlambda][\textlambda][1661,][1666]}

\newcommand{\NIII}{\semiforbiddenEL[N][iii]}
\newcommandx{\NIIIL}[1][1=1750]{\semiforbiddenEL[N][iii][\textlambda][#1]}
\newcommand{\NIIIall}{\semiforbiddenEL[N][iii][\textlambda][\textlambda][1747--][1754]}

\newcommandx{\CIII}{\semiforbiddenEL[C][iii]}
\newcommandx{\CIIIL}[1][1=1909]{\semiforbiddenEL[C][iii][\textlambda][#1]}
\newcommand{\CIIIall}{\semiforbiddenEL[C][iii][\textlambda][\textlambda][1907,][1909]}

\newcommand{\NeIV}{\forbiddenEL[Ne][iv]}
\newcommandx{\NeIVL}[1][1=2424]{\forbiddenEL[Ne][iv][\textlambda][#1]}
\newcommand{\NeIVall}{\forbiddenEL[Ne][iv][\textlambda][\textlambda][2422,][2424]}

\newcommand{\MgII}{\permittedEL[Mg][ii]}
\newcommandx{\MgIIL}[1][1=2803]{\permittedEL[Mg][ii][\textlambda][#1]}
\newcommand{\MgIIall}{\permittedEL[Mg][ii][\textlambda][\textlambda][2796,][2803]}

\newcommand{\NeV}{\forbiddenEL[Ne][v]}
\newcommandx{\NeVL}[1][1=3426]{\forbiddenEL[Ne][v][\textlambda][#1]}
\newcommand{\NeVall}{\forbiddenEL[Ne][v][\textlambda][\textlambda][3346,][3426]}

\newcommand{\OII}{\forbiddenEL[O][ii]}
\newcommandx{\OIIL}[1][1=3727]{\forbiddenEL[O][ii][\textlambda][#1]}
\newcommand{\OIIall}{\forbiddenEL[O][ii][\textlambda][\textlambda][3726,][3729]}

\newcommand{\NeIII}{\forbiddenEL[Ne][iii]}
\newcommandx{\NeIIIL}[1][1=3869]{\forbiddenEL[Ne][iii][\textlambda][#1]}
\newcommand{\NeIIIall}{\forbiddenEL[Ne][iii][\textlambda][\textlambda][3869,][3967]}

\newcommand{\OIII}{\forbiddenEL[O][iii]}
\newcommandx{\OIIIL}[1][1=5007]{\forbiddenEL[O][iii][\textlambda][#1]}
\newcommand{\OIIIall}{\forbiddenEL[O][iii][\textlambda][\textlambda][4959,][5007]}

\newcommandx{\NIL}[1][1=5200]{\forbiddenEL[N][i][\textlambda][#1]}
\newcommand{\NIall}{\forbiddenEL[N][i][\textlambda][\textlambda][5198,][5200]}

\newcommand{\OI}{\forbiddenEL[O][i]}
\newcommandx{\OIL}[1][1=6300]{\forbiddenEL[O][i][\textlambda][#1]}
\newcommand{\OIall}{\forbiddenEL[O][i][\textlambda][\textlambda][6300,][6364]}

\newcommand{\HeI}{\permittedEL[He][i]}
\newcommandx{\HeIL}[1][1=5875]{\permittedEL[He][i][\textlambda][#1]}

\newcommand{\NII}{\forbiddenEL[N][ii]}
\newcommandx{\NIIL}[1][1=6583]{\forbiddenEL[N][ii][\textlambda][#1]}
\newcommand{\NIIall}{\forbiddenEL[N][ii][\textlambda][\textlambda][6548,][6583]}

\newcommand{\SII}{\forbiddenEL[S][ii]}
\newcommand{\SIIL}[1][1=6716]{\forbiddenEL[S][ii][\textlambda][#1]}
\newcommand{\SIIall}{\forbiddenEL[S][ii][\textlambda][\textlambda][6716,][6731]}

\newcommandx{\OIIAuL}[1][1=7325]{\forbiddenEL[O][ii][\textlambda][#1]}
\newcommand{\OIIAuall}{\forbiddenEL[O][ii][\textlambda][\textlambda][7319--][7331]}

\newcommandx{\CIIFIRL}{\forbiddenEL[C][ii][\textlambda][158\,\mum]}

\newcommand{\hda}{\ensuremath{\mathrm{H\text{\textdelta}_A}}\xspace}
\newcommand{\hga}{\ensuremath{\mathrm{H\text{\textgamma}_A}}\xspace}

\title[A 25-kpc nebula around a normal galaxy]{JADES and SAPPHIRES: Galaxy Metamorphosis Amidst a Huge, Luminous Emission-line Region}

\author[\sendemail{francesco.deugenio@gmail.com}{Questions about your JADES paper about G. Mahler}{Greetings Francesco,\%0A\%0Ahow are things this time of the year? I wanted to ask a quick question about this sick object, if I may (great acronym, by the way -- must have been a 'titanic' effort, wink wink).\%0AThe thing is, that ...\%0A\%0ARegards,\%0A}{F. D'Eugenio}~et al.]{\parbox{\textwidth}{
\orcidsymb{Francesco D'Eugenio}{0000-0003-2388-8172}$^{\hyperlink{aff1}{1},\hyperlink{aff2}{2}}$\thanks{E-mail: francesco.deugenio@gmail.com},
\orcidsymb{Jakob M.~Helton}{0000-0003-4337-6211}$^{\hyperlink{aff3}{3}}$,
\orcidsymb{Kevin Hainline}{ 0000-0003-4565-8239}$^{\hyperlink{aff3}{3}}$,
\orcidsymb{Fengwu Sun}{0000-0002-4622-6617}$^{\hyperlink{aff4}{4}}$,
\orcidsymb{Roberto Maiolino}{0000-0002-4985-3819}$^{\hyperlink{aff1}{1},\hyperlink{aff2}{2},\hyperlink{aff5}{5}}$,
\orcidsymb{Pablo G. P{\'e}rez-Gonz{\'a}lez}{0000-0003-4528-5639}$^{\hyperlink{aff6}{6}}$,
\orcidsymb{Ignas Juod{\v z}balis}{0009-0003-7423-8660}$^{\hyperlink{aff1}{1},\hyperlink{aff2}{2}}$,
\orcidsymb{Santiago Arribas}{0000-0001-7997-1640}$^{\hyperlink{aff6}{6}}$,
\orcidsymb{Andrew J. Bunker}{0000-0002-8651-9879}$^{\hyperlink{aff7}{7}}$,
\orcidsymb{Stefano Carniani}{0000-0002-6719-380X}$^{\hyperlink{aff8}{8}}$,
\orcidsymb{Emma Curtis-Lake}{0000-0002-9551-0534}$^{\hyperlink{aff9}{9}}$,
\orcidsymb{Eiichi Egami}{0000-0003-1344-9475}$^{\hyperlink{aff3}{3}}$,
\orcidsymb{Daniel J.~Eisenstein}{0000-0002-2929-3121}$^{\hyperlink{aff4}{4}}$,
\orcidsymb{Benjamin D.~Johnson}{0000-0002-9280-7594}$^{\hyperlink{aff4}{4}}$,
\orcidsymb{Brant Robertson}{0000-0002-4271-0364}$^{\hyperlink{aff10}{10}}$,
\orcidsymb{Sandro Tacchella}{0000-0002-8224-4505}$^{\hyperlink{aff1}{1},\hyperlink{aff2}{2}}$,
\orcidsymb{Christopher N.~A. Willmer}{0000-0001-9262-9997}$^{\hyperlink{aff3}{3}}$,
\orcidsymb{Chris~Willott}{0000-0002-4201-7367}$^{\hyperlink{aff11}{11}}$,
\orcidsymb{William M. Baker}{0000-0003-0215-1104}$^{\hyperlink{aff12}{12}}$,
\orcidsymb{A.~Lola Danhaive}{0000-0002-9708-9958}$^{\hyperlink{aff1}{1},\hyperlink{aff2}{2}}$,
\orcidsymb{Qiao Duan}{0009-0009-8105-4564}$^{\hyperlink{aff1}{1},\hyperlink{aff2}{2}}$,
\orcidsymb{Yoshinobu Fudamoto}{0000-0001-7440-8832}$^{\hyperlink{aff13}{13},\hyperlink{aff3}{3}}$,
\orcidsymb{Gareth C.~Jones}{0000-0002-0267-9024}$^{\hyperlink{aff1}{1},\hyperlink{aff2}{2}}$,
\orcidsymb{Xiaojing Lin}{0000-0001-6052-4234}$^{\hyperlink{aff3}{3}}$,
\orcidsymb{Weizhe Liu \begin{CJK}{UTF8}{gbsn}(刘伟哲)\end{CJK}}{0000-0003-3762-7344}$^{\hyperlink{aff3}{3}}$,
\orcidsymb{Michele Perna}{0000-0002-0362-5941}$^{\hyperlink{aff6}{6}}$,
\orcidsymb{D\'avid Pusk\'as}{0000-0001-8630-2031}$^{\hyperlink{aff1}{1},\hyperlink{aff2}{2}}$,
\orcidsymb{Pierluigi Rinaldi}{0000-0002-5104-8245}$^{\hyperlink{aff3}{3}}$,
\orcidsymb{Jan Scholtz}{0000-0001-6010-6809}$^{\hyperlink{aff1}{1},\hyperlink{aff2}{2}}$,
\orcidsymb{Yang Sun}{0000-0001-6561-9443}$^{\hyperlink{aff3}{3}}$,
\orcidsymb{James A.~A. Trussler}{0000-0002-9081-2111}$^{\hyperlink{aff4}{4}}$,
\orcidsymb{Hannah \"Ubler}{0000-0003-4891-0794}$^{\hyperlink{aff14}{14}}$,
\orcidsymb{Giacomo Venturi}{0000-0001-8349-3055}$^{\hyperlink{aff8}{8}}$,
\orcidsymb{Christina C.\ Williams}{0000-0003-2919-7495}$^{\hyperlink{aff15}{15}}$
and \orcidsymb{Yongda Zhu}{0000-0003-3307-7525}$^{\hyperlink{aff3}{3}}$
}\vspace{0.4cm}
\\
\parbox{\textwidth}{
\hypertarget{aff1}{$^{1}$}Kavli Institute for Cosmology, University of Cambridge, Madingley Road, Cambridge, CB3 0HA, United Kingdom\\
\hypertarget{aff2}{$^{2}$}Cavendish Laboratory - Astrophysics Group, University of Cambridge, 19 JJ Thomson Avenue, Cambridge, CB3 0HE, United Kingdom\\
\hypertarget{aff3}{$^{3}$}Steward Observatory, University of Arizona, 933 N. Cherry Ave., Tucson, AZ, 85721, USA\\
\hypertarget{aff4}{$^4$}Center for Astrophysics $|$ Harvard \& Smithsonian, 60 Garden St., Cambridge MA 02138 USA \\
\hypertarget{aff5}{$^{5}$}Department of Physics and Astronomy, University College London, Gower Street, London WC1E 6BT, UK\\
\hypertarget{aff6}{$^{6}$}Centro de Astrobiolog\'ia (CAB), CSIC–INTA, Cra. de Ajalvir Km.~4, 28850 - Torrej\'on de Ardoz, Madrid, Spain\\
\hypertarget{aff7}{$^{7}$}Department of Physics, University of Oxford, Denys Wilkinson Building, Keble Road, Oxford OX1 3RH, UK\\
\hypertarget{aff8}{$^{8}$}Scuola Normale Superiore, Piazza dei Cavalieri 7, I-56126 Pisa, Italy\\
\hypertarget{aff9}{$^{9}$}Centre for Astrophysics Research, Department of Physics, Astronomy and Mathematics, University of Hertfordshire, Hatfield AL10 9AB, UK\\
\hypertarget{aff10}{$^{10}$}Department of Astronomy and Astrophysics University of California, Santa Cruz, 1156 High Street, Santa Cruz CA 96054, USA\\
\hypertarget{aff11}{$^{11}$}NRC Herzberg, 5071 West Saanich Rd, Victoria, BC V9E 2E7, Canada\\
\hypertarget{aff12}{$^{12}$}DARK, Niels Bohr Institute, University of Copenhagen, Jagtvej 155A, DK-2200 Copenhagen, Denmark\\
\hypertarget{aff13}{$^{13}$}Center for Frontier Science, Chiba University, 1-33 Yayoi-cho, Inage-ku, Chiba 263-8522, Japan\\
\hypertarget{aff14}{$^{14}$}Max-Planck-Institut f\"ur extraterrestrische Physik (MPE), Gie{\ss}enbachstra{\ss}e 1, 85748 Garching, Germany\\
\hypertarget{aff15}{$^{15}$}NSF National Optical-Infrared Astronomy Research Laboratory, 950 North Cherry Avenue, Tucson, AZ 85719, USA
}
}

\date{Submitted to MNRAS}

\pubyear{2025}

\begin{document}
\label{firstpage}
\pagerange{\pageref{firstpage}--\pageref{lastpage}}
\maketitle

\begin{abstract}
We report the discovery of a remarkably large and luminous line-emitting nebula extending on either side of the Balmer-break galaxy \target at $z=5.89$, detected with \jwst/NIRCam imaging in \OIIIall and \Halpha and spectroscopically confirmed with NIRCam/WFSS, thanks to the pure-parallel programme SAPPHIRES. The end-to-end velocity offset is $\Delta v=830\pm130~\kms$.
Nebulae with such large sizes and high luminosities (25-pkpc diameter, $L_{\OIII}=1.2\times10^{10}~\mathrm{L_\odot}$) are
routinely observed around bright quasars, unlike \target.
With a stellar mass of $10^{10.1}~\Msun$, this galaxy is at the knee of the mass function at $z=6$. Its star formation rate declined for some time (10–100 Myr prior to observation), followed by a recent (10 Myr) upturn.
This system is part of a candidate large-scale galaxy overdensity, with an excess of Balmer-break galaxies compared to the field (3~\textsigma). 
We discuss the possible origin of this nebula as material from a merger or gas expelled by an active galactic nucleus (AGN). The symmetry of the nebula, its bubble-like morphology, kinematics, high luminosity, and the extremely high equivalent width of \OIII together favour the AGN interpretation.
Intriguingly, there may be a physical connection between the presence of such a
large, luminous nebula and the possible metamorphosis of the central galaxy
towards quenching.
\end{abstract}


\section{Introduction}

No known physical process other than accreting supermassive black holes (SMBHs) can release sufficient energy
to curtail star formation in massive, central galaxies \citep{silk+rees1998,haehnelt+1998,binney2004}.
Theoretical models and numerical simulations demonstrate that without this source of feedback, violent starbursts would bring forth
a host of titanic galaxies which defy observations \citep[stellar mass $\mstar\sim10^{13}~\Msun$;][]{bower+2012}. Yet the precise mechanisms by which active galactic nuclei (AGNs) shape the star-formation history (SFH) of galaxies are still unclear.
Numerical simulations \citep[e.g.,][]{piotrowska+2022} and observations \citep{terrazas+2017,bluck+2022,bluck+2023} suggest that the protracted absence of star formation in galaxies (quiescence) is best predicted by SMBH mass, which we can interpret as a
`calorimeter' measuring time-integrated feedback from AGNs. Together with evidence of a systematic
increase in stellar metallicity from star-forming to quiescent galaxies \citep{peng+2015,trussler+2020,looser+2024b,baker+2024a}, the emerging scenario is that the energy
released by SMBH accretion accumulates in the halo, which stops the cold gas accretion onto
the central galaxy and thus prevents star formation (preventive feedback scenario).

This coherent picture has been recently challenged by new \jwst observations of massive, quiescent galaxies at redshifts $z=3\text{--}7$ \citep{carnall+2023b,carnall+2024,weibel+2024b}.
The key challenges in the preventive-feedback scenario are the high number density of these quiescent systems
\citep{carnall+2023a,valentino+2023,long+2024,baker+2024,nanayakkara+2024a,nanayakkara+2024b} and their rapid formation and quenching
\citep{carnall+2023b,carnall+2024,glazebrook+2024,degraaff+2024,weibel+2024b,turner+2025}.
By redshift $z=3$, there is already evidence for the existence of quiescent, gas-depleted systems \citep{scholtz+2024}.
While some models are more successful than others \citep{remus+2023,lagos+2024,rennehan+2023},
some form of more powerful or more efficient AGN feedback seems required to match the 
observations \citep{xie+2024}.
Ejective feedback may play an important role, as evidenced by direct
observations of neutral-gas outflows with high mass loading
\citep[$z=2.5\text{--}3$;][]{belli+2024,deugenio+2024a,wu+2025}, with an incidence of 30--40
per~cent in the massive-galaxy population at $z=1\text{--}2.5$ \citep{davies+2024}.
Where these quiescent galaxies with outflows have deep ALMA observations, no
significant molecular gas is detected \citep[with a molecular-gas to stellar mass
fraction of less than 3 percent;][]{scholtz+2024}, consistent with rapid gas 
removal (although gas consumption with no refuelling would also explain the lack
of gas). Intriguingly, one of these outflows is observed in a galaxy without
evidence of AGN \citep{valentino+2025}.

Though alluring, rapid quenching scenarios raise the issue of what are the
star-forming progenitors of these early quiescent systems, because
at $z=7\text{--}11$, compact starbursts with star-formation rates (SFR) of
several hundred \Msun~\peryr should be readily observable by \jwst
\citep[e.g.,][]{weibel+2024b,turner+2025}, but currently remain unseen. At first glance, a 
captivating explanation is that these missing progenitors may be heavily dust obscured as the compact size of the quiescent galaxies and their high metal content are 
both conducive to severe dust attenuation. But even this explanation is not
completely satisfactory: where are the transitional systems between the dust-obscured,
metal-rich progenitor galaxies and their (relatively) dust-free, quiescent descendants?

A possible solution to this conundrum is that during the transition phase, AGNs may
have a high duty cycle, such that most progenitors are AGN hosts. Supporting
evidence for this scenario comes from the discovery that some faint-quasar host galaxies have
post-starburst spectral features, indicating a protracted lack of star formation
\citep{onoue+2024}, but more statistics are needed to confirm this scenario. A number of transitioning progenitors may be hidden by dust-obscured AGNs 
\citep{ross+2015,banerji+2015,hamann+2017,wang+2024b}, though there are claims that these systems may harbour ongoing star
formation \citep{killi+2024}.
Alternatively, the transition from dust-obscured star formation may be extremely rapid,
shortening its visibility time. During this `blowout' phase, one or more AGN events would destroy and/or remove most of the galaxy interstellar medium \citep[ISM;][]{hopkins+2008}. Evidence for such a scenario comes from observations of extended emission-line regions around quasars
\citep[e.g.,][]{perna+2023a,marshall+2023,peng+2024,liu+2024a,liu+2024b,zamora+2024,vayner+2024},
including faint systems \citep{lyu+2024}.
A fraction of this nebular emission could be due to satellite galaxies and their tidally disrupted ISM, because quasars may trace overdensities \citetext{e.g., \citealp{wang+2023a,kashino+2023} -- but see 
\citealp{eilers+2024} for a different view}.
However, a non-negligible part of the ionized 
gas in these extended nebulae may have originally been part of the ISM of the central galaxy before being expelled.
This view is supported by the finding that the observed extended gas emission is metal rich, as suggested by the presence of fast gas outflows in action in quasars and by the very high equivalent width (EW) of the
emission lines which challenges the tidal-disruption and star-formation interpretations \citep{lyu+2024}.

Recent observations of massive, quiescent galaxies at
$z=3\text{--}4$ show the presence of extended and metal-rich gas nebulae
\citep{deugenio+2024a,perez-gonzalez+2024} on scales of the circumgalactic medium (CGM) 
of these compact galaxies. However, it is still unclear if these CGM nebulae are due
to mergers or to outflows \citep{deugenio+2024a}, and if some of them are powered by 
\textit{in-situ} AGNs \citep{perna+2023b}, or by an AGN from the central galaxy
\citep{perez-gonzalez+2024}, which could also be obscured along our line of sight or faded away, leaving a fossil ionized nebula, as seen in \textit{voorwerpen} 
\citep{lintott+2009,keel+2015,finlez+2022,venturi+2023,solimano+2025}. Investigating the physical properties of these 
extended gas nebulae and relating their presence and properties to the star-formation history of their central 
galaxies may hold the key to understanding how SMBHs quench galaxies \citep{nelson+2019}. 
In turn, this may reveal how massive, quiescent galaxies seemingly appear out of nowhere.

In this article, we report the discovery of a candidate `blowout' system at redshift $z=6$.
This galaxy, \target, is found at the centre of a large overdensity of galaxies, and is surrounded by a
remarkably extended and luminous emission-line nebula. After presenting the data (Section~\ref{s.obs}), we
describe the redshift determination of the system (Section~\ref{s.redshift}) and the morphology of the central galaxy (Section~\ref{s.morph}).
Section~\ref{s.sed} presents the spectral energy distribution (SED) of the main components of
the system, while the energetics of the emission-line regions are discussed in Section~\ref{s.eml}.
In Section~\ref{s.environment} we investigate the large-scale environment, before proceeding to interpret
our findings and discuss their implications (Section~\ref{s.disc}). A short summary is provided in
Section~\ref{s.conc}.

Throughout this work, we assume the \citet{planck+2020} \textLambda CDM cosmology, where we calculate a physical
scale of 5.89~kpc~arcsec$^{-1}$ at redshift $z=5.89$. Unless otherwise specified, physical scales are given as proper quantities. All stellar masses are total stellar mass
formed, assuming a \citet{chabrier2003} initial mass function, integrated between 0.1 and
120~\Msun. All magnitudes are in the AB system \citep{oke+gunn1983} and all EWs are in the rest frame, with negative EW corresponding to line emission.

\section{Data}\label{s.obs}

\begin{figure*}
   \includegraphics[width=\textwidth]{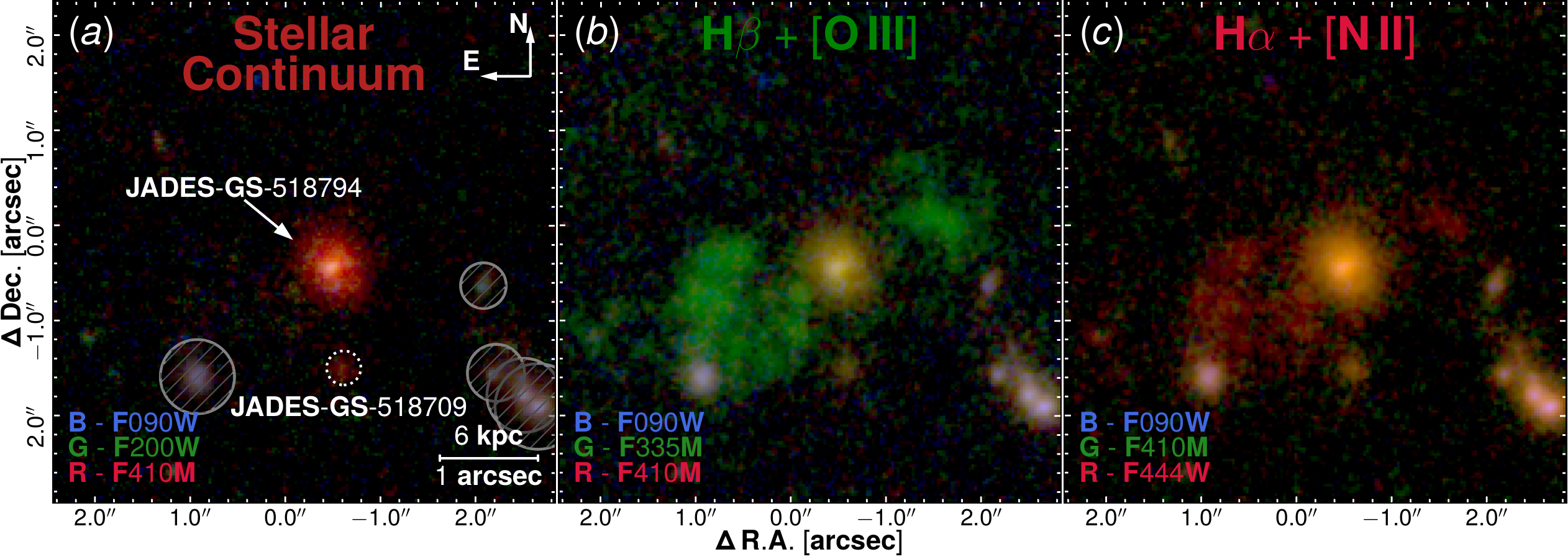}
   {\phantomsubcaption\label{f.img.a}
    \phantomsubcaption\label{f.img.b}
    \phantomsubcaption\label{f.img.c}}
   \caption{\jwst/NIRCAM false-colour images of \target, highlighting continuum emission (panel~\subref{f.img.a}),
   \Hbeta and \OIIIall (captured by F335M; panel~\subref{f.img.b}), and \Halpha and \NIIall (falling in
   F444W; panel~\subref{f.img.c}).
   Hatched sources in panel~\subref{f.img.a} are foreground galaxies (identified via conspicuous
   flux blueward of the \Lyalpha break at $z=5.89$, detected by \hst/ACS). The galaxy \target is located at
   the centre of a 25-kpc diameter nebula, extending on both sides and displaying arc-like shapes,
   reminiscent of shock fronts/hot gas bubbles.
   A possible quiescent satellite is also present (JADES-GS-518709, dotted circle in panel~\subref{f.img.a}, also visible in panels~\subref{f.img.b} and~\subref{f.img.c}).}\label{f.img}
\end{figure*}

Our target galaxy \target (Fig.~\ref{f.img}) is a red, faint galaxy ($F_{F444W} = 0.75\pm0.01$~\textmu Jy, 24.2~mag\footnote{We used a Kron aperture, determined as described in \citet{rieke+2023}. This may incur in contamination from diffuse \Halpha, but the flux from modelling the light profile in F410M (where no contamination from diffuse nebular emission is present) is comparable, at $F_{F410M}=0.67\pm0.01$~\textmu Jy or 24.3~mag. See Section~\ref{s.morph}.})
at R.A. 53.00804,
Dec. $-27.85477$, located in the GOODS South (GOODS-S) cosmological field
\citetext{Great Observatories Origins Deep Survey, \citealp{giavalisco_goods_2004}}. The source was first detected by the \textit{Spitzer}
SIMPLE Survey in IRAC ch1 and ch2 \citep[][SIMPLE~ID~26553; R.A. 53.007931, Dec.$-27.854734$; separation
0.4~arcsec]{damen+2011}. The IRAC ch2 (4.5~\mum) flux is $F_{ch2} = 1.7\pm0.2$~\textmu Jy, higher than the \jwst value by a factor of two. This flux mismatch and the centroid offset
are possibly due to source confusion between the galaxy, the surrounding 
nebula, and nearby foreground galaxies. This is exacerbated by the much larger point spread function (PSF) of IRAC ch2 vs NIRCam F444W \citetext{full-width half-maximum, FWHM=1.9 arcsec \citealp{damen+2011} vs 0.14 arcsec \citealp{ji+2024b}}. No detection at redder wavelengths is
reported, so from here on we focus on the new data from \jwst.
Our NIRCam observations come from the Guaranteed Time Observations 
\citep[JADES, \jwst Advanced Deep Extragalactic Survey;][]{eisenstein+2023a},
obtained as part of program ID (PID)~1286 (PI N.~L\"utzgendorf), and consists of 5.67--11.34~ks integrations 
using various wide- and medium-band filters, as detailed in Table~\ref{t.imaging}. We supplement the JADES NIRCam observations with legacy \hst/ACS imaging at shorter wavelengths from the Hubble Legacy Field data \citep{whitaker+2019}. 
The NIRCam data reduction follows the approach presented in \citet{rieke+2023}, \citet{eisenstein+2023b}, \citet{robertson+2024} and \citet{deugenio+2024}.
Improvements with respect to the procedures highlighted in these papers will be presented in
a future data-release article (JADES collaboration, in prep.).

For the main galaxy \target, we also use mid-infrared (MIR) photometry from 
\textit{Spitzer}/IRAC and MIPS and from \jwst/MIRI. For \textit{Spitzer}, we 
retrieve images from the Rainbow database obtained with IRAC ch3 and ch4
\citetext{5.8 and 8~\mum, \citealp{perez-gonzalez+2008,guo+2012}} and with
MIPS 24~\mum \citep{perez-gonzalez+2005}.
Since the source is detected at most tentatively, we use photometry extracted from circular apertures centred on the nominal position of the target.
For IRAC ch3 we use apertures of 1.5 and 2-arcsec radius, which give the same 
flux of $24.5\pm0.5$~mag (after applying aperture corrections for point-like 
sources). For the noisier ch4, we use a smaller 0.75-arcsec aperture to derive 
an upper limit of 23.2~mag. Similarly, we find only an upper limit in MIPS
24~\mum. For MIRI, we use data from PID~1180, which was not previously reduced due to a guide-star failure. However, the observations in F1500W show a clear detection at
the location of JADES-GS-518794.

Photometric redshifts $z_\mathrm{phot}$ were obtained using \eazy 
\citep{brammer+2008}, using photometry from JADES~DR3 \citep{deugenio+2024} 
and the procedure described in \citet{hainline+2024}. Throughout this paper, we adopt the \verb|z_a| \eazy photometric redshifts described in \citet{hainline+2024}, and the uncertainties are defined as (\verb|u68-l68|)/2. For the new region of JADES where \target lies, these measurements will be released in the future (Hainline, K., in prep.).

\begin{table}
    \setlength{\tabcolsep}{4pt}
    \caption{Summary of the new NIRCam observations (PID~1286, Obs~6) used to find and study \target.
    An in-depth presentation of data from PID~1286 will be presented in JADES Collaboration (in~prep.).}
    \label{t.imaging}
    \begin{tabular}{lclc}
  \toprule
    Filter Name& $t_\mathrm{exp}$ & Filter Name& $t_\mathrm{exp}$ \\
               &        [ks]      &            &        [ks]      \\
  \midrule
    F070W & 5.67  &  F277W & 8.50  \\
    F090W & 11.34 &  F335M & 5.67  \\
    F115W & 11.34 &  F356W & 5.67  \\
    F150W & 8.50  &  F410M & 11.34 \\
    F200W & 5.67  &  F444W & 11.34 \\
  \bottomrule
    \end{tabular}
\end{table}

When characterizing the large-scale environment, photometric redshifts are complemented by slit spectroscopy from the NIRSpec Micro-Shutter Assembly
\citetext{\citealp{ferruit+2022,jakobsen+2022}; obtained by JADES \citealp{eisenstein+2023a}} and by slitless spectroscopy obtained as part of the programmes:
FRESCO \citetext{First Reionization Epoch Spectroscopic COmplete Survey; PID~1895, \citealp{oesch+2023}}, SAPPHIRES (Slitless Areal Pure Parallel HIgh-Redshift Emission Survey; PID~6434, PI~E.~Egami), and PID~4540 (JADES; PI D.~Eisenstein).

NIRCam wide-field slitless spectroscopy (WFSS) of \target was serendipitously obtained through SAPPHIRES. 
The row-direction grism (Grism-R) observation was obtained with their observation number 41 through the F444W filter (3.9--5.0~\mum) with a total on-source integration time of 11.8~hours.
Detailed observation description will be presented by a paper through the SAPPHIRES collaboration (Sun, F.\ et al., in prep).

\section{Redshift determination}\label{s.redshift}

Both \target and the extended region surrounding it  (Fig.~\ref{f.img}) display clear photometric excess in F335M relative to F356W, and in F444W relative to F410M. The magnitude of these differences leaves no doubt about their interpretation as emission lines with high equivalent width. At the same time, the galaxy is a clear F070W dropout and
very faint in F090W, suggesting a redshift around $z\sim6$. Around this redshift, \Hbeta and \OIIIall emission would fall in F335M, while
\Halpha would fall in F444W, redward of F410M.
The estimated photometric redshift of \target is $z_\mathrm{phot}=6.03^{+0.16}_{-0.15}$. The extended region surrounding the galaxy consists of two main regions (hereafter: East and West clouds; Fig.~\ref{f.img.b}--\subref{f.img.c}).
The standard JADES imaging pipeline deblends these two clouds into ten sources. Of these, eight are
consistent with the photometric redshift of
\target, while the two brightest regions are
placed by \eazy at $z_\mathrm{phot}=0.01$, with
a secondary $\chi^2$ minimum at $z_\mathrm{phot}\sim 6$. This redshift mismatch is likely due to the fact that the photometry for these sub-regions is dominated by ionized gas emission, with very little underlying galaxy light, and \eazy cannot reproduce the observed colours with galaxy templates. Indeed, when performing
SED modelling, we infer extremely high EWs, which we discuss further in Section~\ref{s.sed}.
In any case, these two regions at $z_\mathrm{phot}=0.01$ are not 
detected by \hst, and their minimum $\chi^2$ value
is unsatisfactorily large ($\chi^2 = 140$ and 160).
Other nearby objects detected by \hst are clear
foreground galaxies. These have distinctively
bluer colour, no photometric excess in the NIRCam
filters reported above, and reliable $z_\mathrm{phot}$ placing them at various redshifts between 1.5 and 3. These foreground galaxies are marked by
hatched circles in Fig.~\ref{f.img.a}.

We determine the spectroscopic redshift of \target through NIRCam WFSS.
The SAPPHIRES grism spectra were processed through the routine outlined by \citet{sunf+23}, and the codes and calibration data are publicly available\footnote{\url{https://github.com/fengwusun/nircam_grism/}}.
We refer interested readers to the SAPPHIRES early data release article for updated descriptions of NIRCam WFSS calibration and SAPPHIRES data processing \citep{sun+2025}.

\begin{figure*}
\centering
\includegraphics[width=\linewidth]{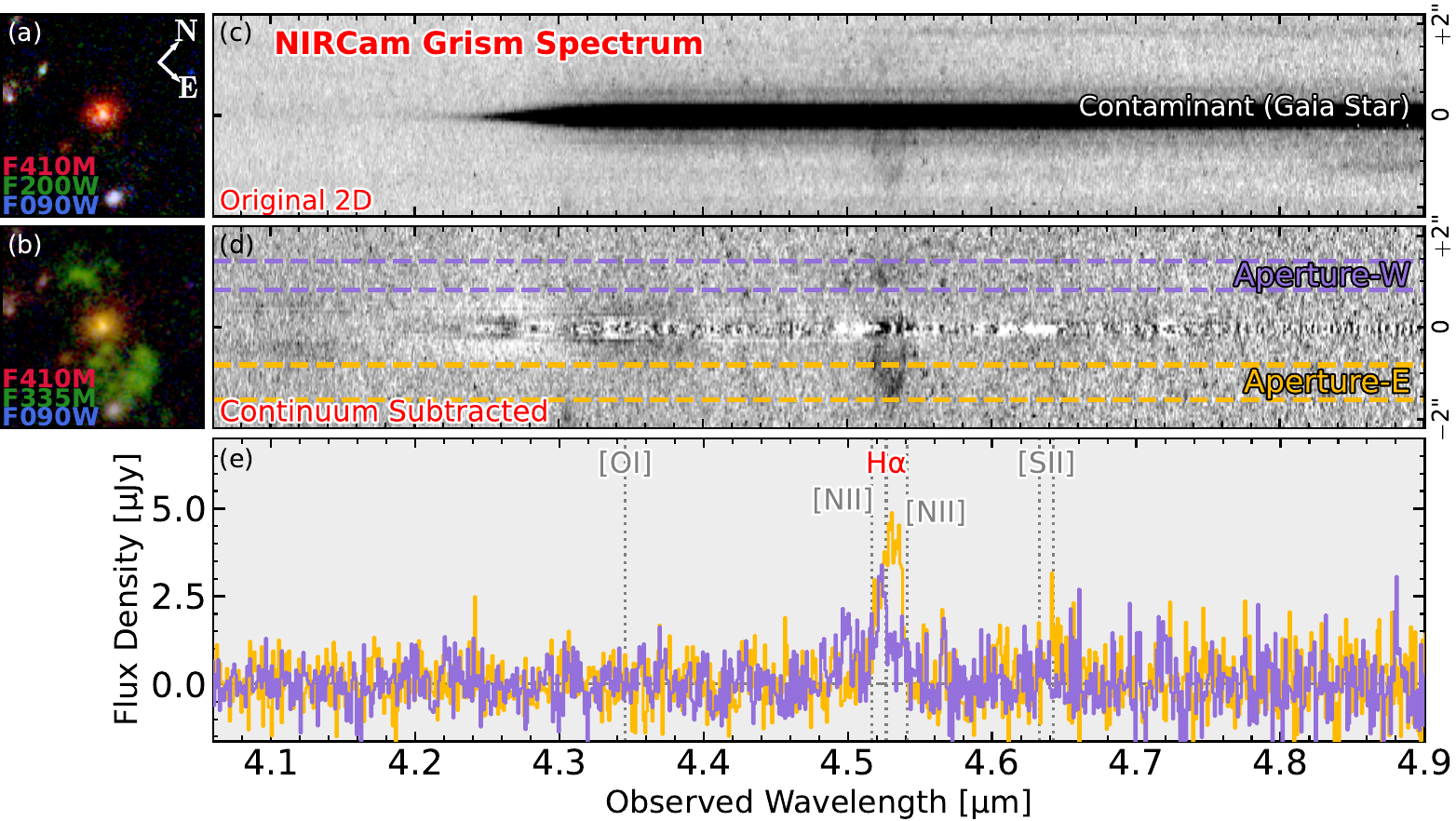}
  {\phantomsubcaption\label{f.grism.a}
   \phantomsubcaption\label{f.grism.b}
   \phantomsubcaption\label{f.grism.c}
   \phantomsubcaption\label{f.grism.d}
   \phantomsubcaption\label{f.grism.e}}
\caption{NIRCam F444W grism spectrum of \target obtained through Cycle-3 PID~6434 (SAPPHIRES).
  Panels~\subref{f.grism.a} and~\subref{f.grism.b} are the same false-colour RGB images as in
  Fig.~\ref{f.img.a} and~\subref{f.img.b}, highlighting respectively the stellar continuum and
  \OIII line emission; the images have been rotated and flipped to align them with the dispersion direction
  of the 2D grism spectrum, shown on the right. Panels~\subref{f.grism.c} shows the 2D grism spectrum;
  the bright continuum is from a contaminating \textit{Gaia} star. Panel~\subref{f.grism.d} shows the
  2D grism spectrum with the continuum subtracted; two boxcar apertures (W/E) for spectral extraction are shown in pink and yellow, respectively, capturing emission from the East and West clouds (see
  Fig.~\ref{f.sed}). Panel~\subref{f.grism.e} shows the 1D grism spectra extracted with the two apertures; for both spectra, we identify the line at $\sim 4.52~\mum$ as \Halpha. Other notable
  emission lines are indicated by vertical dotted lines (grey labels are non-detections).
}\label{f.grism}
\end{figure*}

Fig.~\ref{f.grism} displays the reduced F444W grism spectrum of our target, which covers the spectral region including \Halpha, \NIIall and \SIIall.
\target's spectrum is severely contaminated by the spectrum of the bright \textit{Gaia} star (Gaia DR3 5057502837774749952), which hampers measuring the central galaxy's redshift.
However, diffuse line emission is visible at $\sim4.52$~\mum, which is spatially offset from the bright continuum contaminant. 
We interpret the emission as \Halpha emission at $z\sim5.89$ according to the JADES $z_\mathrm{phot}$. 
The diffuse \Halpha emission is spatially aligned with two line-emitting regions identified in NIRCam imaging data, which are to the west and east of the central galaxy (Fig.~\ref{f.img}).
We therefore extract the spectra for the two regions using boxcar apertures, and the heights are 0.57 and 0.69~arcsec for aperture-W and -E, respectively. 
The continuum is modelled and subtracted through smoothing spline interpolation with line-emitting wavelength (4.49--4.55~\mum) masked.
We fit the emission lines with a Gaussian profile in the continuum-subtracted 1D spectra. 
After considering the spatial offsets of the line-emitting region relative to the central galaxy in both NIRCam image and spectra, we measure spectroscopic redshifts of 5.903 and 5.884 for the West and East clouds, respectively.
The redshift uncertainty is estimated to be $\Delta z = 0.002$, which is dominated by the uncertainty of centroid determination in both imaging and grism data. The redshift offset between the two clouds corresponds to a velocity
offset of $830\pm130~\kms$.
The spectrum of the East clouds appears dominated by \Halpha emission, with no evidence for \NIIall or \SIIall. The spectrum of the West cloud displays a secondary emission-line peak near the
wavelength of \NIIL, reaching 30~percent of the \Halpha peak. However, this emission could also be due to \Halpha, spectrally offset with respect to the main peak due to the spatially extended nature of the cloud and to
the grism degeneracy between spatial offset and spectral shift. Given this 30~percent upper limit and the fainter nature of the West cloud, we can safely conclude that at
wavelengths around 4.5~\mum, the emission is dominated by \Halpha.

\section{Morphological parameters}\label{s.morph}

\begin{figure}
  {\phantomsubcaption\label{f.profs.a}
   \phantomsubcaption\label{f.profs.b}
   \phantomsubcaption\label{f.profs.c}
   \phantomsubcaption\label{f.profs.d}
   \phantomsubcaption\label{f.profs.e}
   \phantomsubcaption\label{f.profs.f}
   \phantomsubcaption\label{f.profs.g}
   \phantomsubcaption\label{f.profs.h}
   \phantomsubcaption\label{f.profs.i}
   \phantomsubcaption\label{f.profs.j}
   \phantomsubcaption\label{f.profs.k}
   \phantomsubcaption\label{f.profs.l}}
  \includegraphics[width=\columnwidth]{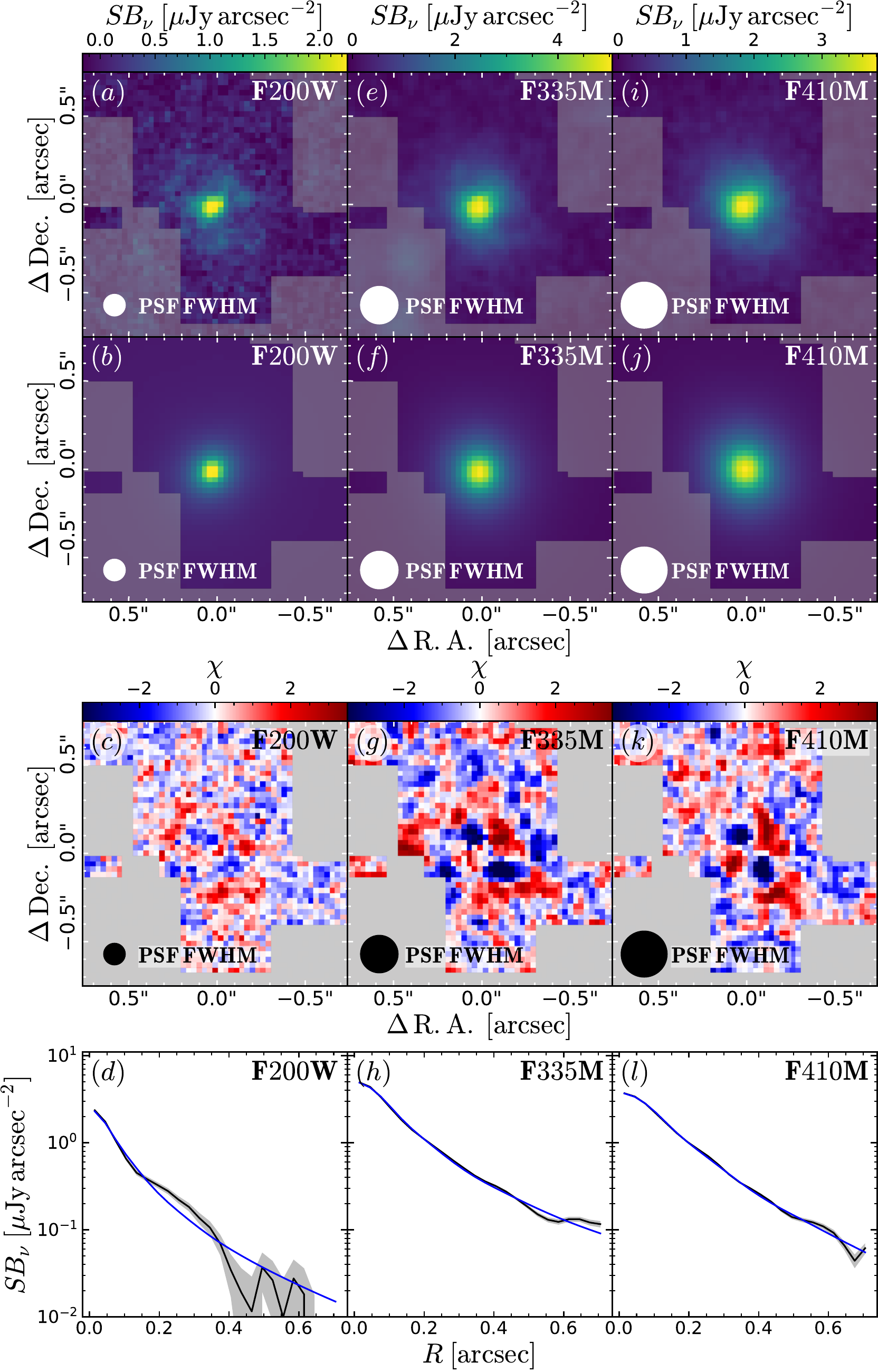}
  \caption{Morphological analysis of \target, with each column showing the results for a different NIRCam filter as indicated. From top to bottom, the rows are the data, fiducial S\'ersic model, $\chi$ residuals, and radial surface-brightness profile. The grey rectangles mask nearby sources that are not included in the fit.}\label{f.profs}
\end{figure}

We use \pysersic \citep{pasha+miller2023} to measure the structural parameters of \target, using a single S\'ersic light profile. We limit our models to F200W, capturing the
rest-frame near-UV blueward of the Balmer break, F335M (tracing \Hbeta and \OIII emission), and
F410M, tracing the stellar continuum in the rest-frame optical. In each band, we use empirical PSFs measured from the
JADES data \citep{ji+2024b}. Nearby and foreground sources are masked (Fig.~\ref{f.profs}). The fiducial model is the maximum \textit{a posteriori} model with a Student's $t$ loss function, while the marginalized posterior probabilities on the free parameters are estimated using the stochastic variational inference method \citep{hoffman+2013}.

A summary of the analysis is presented in Fig~\ref{f.profs}; the marginalized posterior probabilities of the model parameters are
reported in Table~\ref{t.morph}. In all bands, the model prefers a S\'ersic index of 2--3, particularly at the bluest wavelengths. The half-light semi-major axes span 0.21--0.29~arcsec, corresponding to 1.2--1.7~kpc, which is typical for star-forming galaxies at this redshift \citep{ormerod+2023}, but much larger than quiescent galaxies \citetext{by extrapolating measurements at $z\lesssim5$, \citealp{ji+2024c}, or via direct measurements at $z=7.3$, \citealp{weibel+2024b}}.
The galaxy images and S\'ersic index reaching $n=3$ suggest the presence of a compact core 
surrounded by more diffuse emission on galaxy scales, but modelling the galaxy with an additional point source is disfavoured, with the
fiducial point-source flux statistically consistent with 0.

\begin{table}
    \caption{Morphological parameters of the central galaxy \target with \pysersic.
    }\label{t.morph}
    \begin{tabular}{lccc}
  \toprule
    \multirow{2}{*}{Filter}  & Flux & \re & S\'ersic $n$ \\
                             & [\textmu Jy] & [arcsec] & --- \\
  \bottomrule
    F200W        &  $0.19\pm0.02$ & $0.21\pm0.03$ & $2.8\pm0.3$ \\
    F335M        &  $0.91\pm0.02$ & $0.29\pm0.02$ & $3.1\pm0.1$ \\
    F410M        &  $0.65\pm0.01$ & $0.22\pm0.01$ & $2.1\pm0.1$ \\
  \bottomrule
    \end{tabular}
\end{table}

Inspecting the fit residuals, we find evidence of substructures not captured by a single S\'ersic profile. These substructures appear as regions where the residuals have consistently positive residuals (particularly evident in F200W in panel~\ref{f.profs.c}). In
addition, in F200W, the presence of substructures is also evident 
in the excess of the observed surface-brightness profile relative to
the best-fit model (black vs blue lines in panel~\ref{f.profs.d}).
These residual structures appear aligned almost orthogonally to the position angle we infer from the regions with the highest surface brightness. Due to their irregular structures, these residuals suggest some recent merger event, rather than a dynamically relaxed component.

\section{SED Modelling}\label{s.sed}

We model the SED of each source using \prospector \citep{johnson+2021}, following the
setup of \citet[][hereafter \citetalias{tacchella+2022a}]{tacchella+2022a}.
The redshift is free, but we adopt a Gaussian prior with mean $z = 5.89$ and dispersion
$\sigma = 0.01$. Preliminarily, we tested a Gaussian prior with
mean $z=z_\mathrm{phot}$ and dispersion $\sigma = 0.5$, and verified that the resulting posterior probability is statistically consistent with the spectroscopic redshift. We use flat priors on both stellar mass \mstar and metallicity $Z_\star$ (in log space). We
test different SFH parametrizations, but as a fiducial model we use a piecewise constant SFH over ten time
bins (using six time bins gives statistically consistent results). The first three bins are chosen between $t=0\text{--}10$~Myr, $10\text{--}30$~Myr, and
$30\text{--}100$~Myr, with $t$ being the look-back time from the epoch of observation. The remaining
seven bins are logarithmically spaced between $t=100$~Myr and $t=730$~Myr, the look-back time between $z=5.89$ and $z=20$. The SFR is set to 0 at any epoch earlier than $z=20$.
In addition to the total mass formed, this parametrization includes nine free parameters, which are
taken to be the logarithm of the SFR ratio between adjacent time bins. We adopt a probability prior
that favours a rising SFR with time, mirroring the increasing accretion rate of dark matter haloes
with cosmic time \citep[e.g.,][]{wechsler+2002,neistein+dekel2008,dekel+2013,tacchella+2018b}. We use
the implementation presented in \citet{turner+2025} for the massive quiescent galaxy
ZF-UDS-7329 \citep{glazebrook+2024}. Remarkably, this prior yields a better agreement
between the SFH inferred from the $z\sim 3$ observations and direct measurements of the 
SFR of galaxies at earlier epochs \citep{turner+2025}.
The prior is enforced by a Student's $t$ probability distribution on the logarithm of the
ratio of the SFRs between adjacent time bins, with a redshift-dependent mean for each time bin
\citep[Eq.~4 in][]{turner+2025} and with a generous scale parameter of 1, which allows 
$>1$~dex variations in SFR.
We use a variable dust-attenuation law, parametrized as a free power-law modification
index $n$ with respect to the Calzetti curve \citep{calzetti+2000,noll+2009}. This
dust-attenuation index is tied to the strength of the UV bump, following the empirical 
relation by \citet{kriek+conroy2013}. The overall attenuation is parametrized by the
$V$-band equivalent optical depth $\tau_V$, with an additional power-law attenuation
screen towards birth clouds \citep{charlot+fall2000}, assumed to fully enshroud all stars
younger than 10~Myr. We use the \citet{madau+1995} inter-galactic medium absorption
profile.
Stars with ages younger than 10~Myr are associated with nebular emission, which is 
parametrized by the ionization parameter $U$ and by the gas metallicity $Z_\mathrm{gas}$.
For each guess of these two model parameters, the strength of the nebular emission is 
obtained from pre-computed \cloudy models \citep{ferland+2017}, using the setup presented
in \citet{byler+2017}. Each parameter is assigned a prior with varying degrees of information, which we summarize in Table~\ref{t.prosp}.
By adopting this complex model, we do not claim nor
hope to meaningfully constrain all 17 free parameters, particularly given that a fit to nine photometric datapoints alone cannot break the degeneracy between stellar age, metallicity and dust
attenuation \citep{trager+1998}. However, this setup enables us to 
obtain realistic uncertainties and degeneracies in the most relevant parameters we are 
interested in: \mstar and recent trends in the SFR.

\subsection{Central galaxy}\label{s.sed.ss.central}

The marginalized posterior probabilities of the model parameters for \target are reported in 
Table~\ref{t.prosp}.
When using as input the photometric redshift from \eazy,
the posterior probability on the \prospector redshift is fully consistent with the spectroscopic redshift of the extended nebula (Section~\ref{s.redshift}). Therefore, to 
reduce the uncertainties, we use as default the much narrower probability prior from SAPPHIRES. In this case,
the prior is more informative than the photometry, so the
posterior is identical to the prior probability.

\begin{figure*}
   \includegraphics[width=0.91\textwidth]{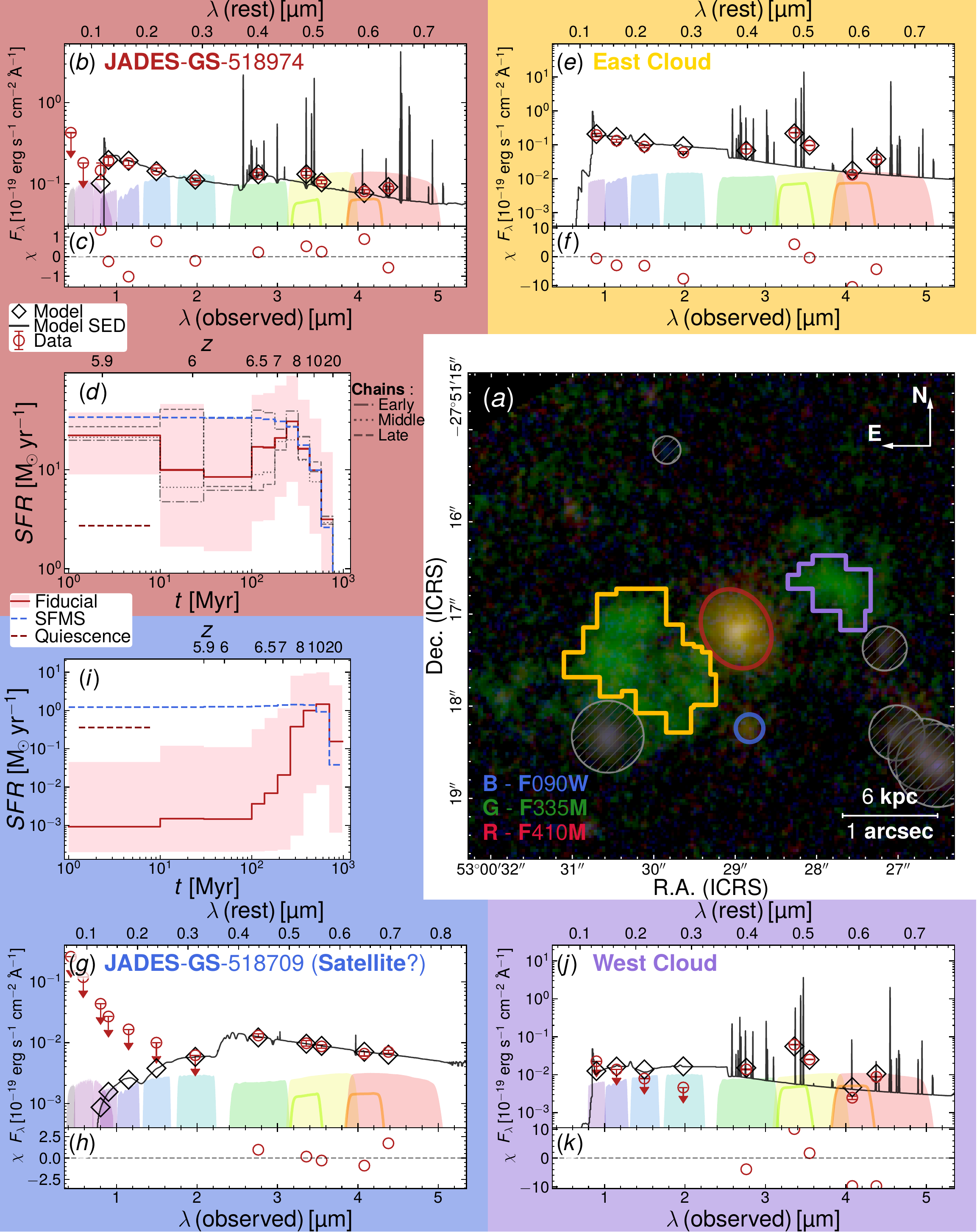}
   {\phantomsubcaption\label{f.sed.a}
    \phantomsubcaption\label{f.sed.b}
    \phantomsubcaption\label{f.sed.c}
    \phantomsubcaption\label{f.sed.d}
    \phantomsubcaption\label{f.sed.e}
    \phantomsubcaption\label{f.sed.f}
    \phantomsubcaption\label{f.sed.g}
    \phantomsubcaption\label{f.sed.h}
    \phantomsubcaption\label{f.sed.i}
    \phantomsubcaption\label{f.sed.j}
    \phantomsubcaption\label{f.sed.k}}
   \caption{Apertures (panel~\subref{f.sed.a}) and \prospector SED models (panels~\subref{f.sed.b}--\subref{f.sed.k}). Observed SEDs (red circles)
   and fiducial models (black diamonds) are accompanied by the model SED
   (black line) and residuals $\chi$ (adjacent bottom panels). The 
   \hst/ACS and \jwst/NIRCam filters are highlighted by the shaded regions, while thick solid lines are NIRCam medium-band filters. For the main target
   \target and the candidate satellite we also show the SFH 
   (panels~\subref{f.sed.d} and~\subref{f.sed.i}, with the star-forming main sequence of \citet{popesso+2023} shown by the blue dashed line). These two galaxies display 
   a clear break at 2.5~\mum, corresponding to a Balmer break at $z\approx 6$. 
   Both SFHs display the early rise due to the prior; the satellite then 
   quenches at $z=7\text{--}8$ and becomes quiescent. \target also displays a 
   downturn in SFR around 10--100~Myr before observation, while a recent upturn in the last 10~Myr could be due to the photometric excess, consistent
   with both a resurrection of the SFR or strong AGN.
   The downturn in SFR is robust and driven by the observed Balmer break; the 
   large uncertainties about the SFR in the range 10--100~Myr reflect the inability of the data to distinguish scenarios of early vs late downturn (see median of individual chains in grey). The line-emitting clouds
   (panels~\subref{f.sed.e}--\subref{f.sed.f} and 
   \subref{f.sed.j}--\subref{f.sed.k}) are poorly fit by \prospector, 
   supporting their interpretation as gas clouds and not tidally disrupted
   satellites.
  }\label{f.sed}  
\end{figure*}

The observed SED and fiducial model are in Fig.~\ref{f.sed.b}, with the fit residuals in panel~\subref{f.sed.c}.
The galaxy SED displays a relatively strong Balmer break \citep[$D(Balmer)=2.24^{+0.16}_{-0.18}$, using the definition of][]{binggeli+2019}, with a mass-weighted stellar
age of $230\pm70$~Myr, intermediate between recently quenched galaxies \citetext{e.g., JADES-GS-z7-01-QU, \citealp{looser+2024}; \citealp{strait+2023}, 
\citealp{baker+2025}} and
quiescent galaxies \citetext{like RUBIES-UDS-QG-z7; \citealp{weibel+2024b}}.
The stellar mass 
is of order $\mstar = 10^{10}~\Msun$, similar to that of RUBIES-UDS-QG-z7 and 
much larger than JADES-GS-z7-01-QU.

The SFH (Fig.~\ref{f.sed.d}) displays a rapid rise at early times, but this is driven
entirely by the adopted prior. At later times the SFR declines (in the last 100--10~Myr),
followed by an upturn in the most recent 10~Myr.  The inference of this downturn is driven
by the large flux difference between F200W and F277W, the bands bracketing the Balmer 
break. \prospector considers the flux excess of F277W with respect to F200W to be too 
large to be due to emission lines (e.g., \OIIall, \NeIIIall), and fits this excess as a stellar Balmer break (Fig.~\ref{f.sed.b}). Indeed, among the 130,000 models explored by prospector
during the inference, the model with the strongest \OIIall emission ($EW(\OII) = -272$~\AA)
still requires a noticeable Balmer break $D(Balmer)=1.99$. Besides this relatively high value, this particular SED model is highly disfavoured
by the data, with a difference in log evidence of 11 relative to the maximum \textit{a-posteriori}
model. We can also rule out strong dust
attenuation causing the flux drop between F277W and F200W, because the continuum appears to be rather blue in the rest-frame UV. Notice that the model
prefers a Balmer break despite the rising SFH prior, which favours high SFR at recent
times and hence is biased against a Balmer break.
Rare scenarios like strong shocks could power extremely bright \OIIall emission \citep{helton+2021}, but these cases are extremely rare around non-quasar galaxies \citep{deugenio+2025}.

The scatter about the fiducial SFH is particularly large
in the time window between 10 and 100~Myr prior to observation, calling into question the
robustness of the claimed downturn. But this large scatter does not take into account that
the SFR values in different time bins are strongly anti-correlated with each other, due to
the inability of the data to break the degeneracy between early vs late declines in the SFR. This anti-correlation arises from the fact that any over-estimate of the SFR at early times leads to an under-estimate at later times, while keeping the strength of the Balmer
break appoximately the same.
To illustrate this point, in Fig.~\ref{f.sed.d} we also show the median SFH of all the
chains whose SFR peaks earlier than 100~Myr (Early, dot-dashed grey line),
between 30--100~Myr (Middle, dotted grey line), and between 10--30~Myr (Late, dashed line).
These three SFHs underscore that there are no acceptable solutions where the galaxy kept forming
stars at a constant rate throughout the period 10--100 Myr; after all, this would not
be consistent with the Balmer break. Hence the inferred downturn in SFR is robust, even
though with photometry alone we cannot time this event precisely.

At more recent times $t<10$~Myr, there is a recent upturn in SFR which is needed to
explain the flux excess in F335M and F444W with respect to 
F356W and F410M, respectively. This photometric excess is interpreted by \prospector as 
being due to recent star formation,
causing bright \Hbeta and \OIIIall in F335M and \Halpha in F444W. However, this 
strong line excess could also be caused by a type-1 AGN, hence the recent 10-Myr upturn 
in SFR cannot be tested without spectroscopy.

To understand the robustness of these results 
 against the adopted SFH prior, we also considered the `continuity' 
prior, with the same time bins as the fiducial model, but with the prior probability on 
the log SFR ratios centred at 0 with a scale of 0.3 \citep{leja+2019}; and the `bursty' 
prior, same as the continuity prior but with a scale of 1 \citep{tacchella+2022a}. The `continuity' and `bursty'
priors implicitly assume a constant SFH across all of the time bins.
The model inference with either of these priors yields consistent redshifts, stellar 
mass and recent shape of the SFH. The major difference is in the mass-weighted stellar
age, where the rising SFH prior results in younger age (Table~\ref{t.prosp}) by 
suppressing the earliest stages of star formation (yet the age difference between the fiducial and alternative models is not statistically significant, only 1.5~\textsigma).

\begin{table*}
    \begin{center}
    \caption{Summary of the model parameters, prior and posterior probabilities for the
    central galaxy \target from \prospector.
    }\label{t.prosp}
    \setlength{\tabcolsep}{4pt}
    \begin{tabular}{lllccc}
  \toprule
   Parameter & Description & Prior & Fiducial & Continuity & Bursty \\
   (1)       & (2)         & (3)   & (4)      & (5)        & (6)    \\
   \midrule
   $z_\mathrm{obs}$       & redshift                                                                                                 & $\mathcal{N}(z_\mathrm{spec}, 0.01)$ & $5.89^{+0.01}_{-0.01}$  & $5.89^{+0.01}_{-0.01}$  & $5.89^{+0.01}_{-0.01}$  \\
   $\log \mstar [\Msun]$  & total stellar mass formed                                                                                & $\mathcal{U}(6, 13)$                 & $10.1^{+0.1}_{-0.2}$ & $10.2^{+0.1}_{-0.1}$ & $10.1^{+0.1}_{-0.1}$ \\
   $\log Z [\Zsun]$       & stellar metallicity                                                                                      & $\mathcal{U}(-2, 0.19)$              & $-0.7^{+0.6}_{-0.5}$ & $-0.7^{+0.6}_{-0.6}$ & $-0.5^{+0.5}_{-0.5}$ \\
   $\log SFR$ ratios      & ratio of the $\log SFR$ between adjacent bins of the non-parametric SFH                                  & $\mathcal{T}(\Xi_i, 1, 2)^a$         & ---                  &        ---           & ---                  \\
   $n$                    & power-law modifier of the dust curve \citepalias[][their eq.~5]{tacchella+2022a}                         & $\mathcal{U}(-1.0,0.2)$              & $-0.0^{+0.2}_{-0.2}$ & $-0.1^{+0.2}_{-0.2}$ & $-0.2^{+0.3}_{-0.3}$ \\
   $\tau_V$               & optical depth of the diffuse dust \citepalias[][their eq.~5]{tacchella+2022a}                            & $\mathcal{G}(0.3,1;0,2)$             & $0.5^{+0.2}_{-0.2}$  & $0.5^{+0.2}_{-0.2}$ & $0.3^{+0.2}_{-0.1}$   \\
   $\mu$                  & ratio between the optical depth of birth clouds and $\tau_V$ \citepalias[][their eq.~4]{tacchella+2022a} & $\mathcal{G}(1,0.3;0,2)$             & $1.0^{+0.4}_{-0.4}$  & $1.0^{+0.3}_{-0.3}$ & $1.0^{+0.4}_{-0.3}$   \\
   $\log Z_\mathrm{gas} [\Zsun]$ & metallicity of the star-forming gas & $\mathcal{U}(-2, 0.19)$                                     & $0.0^{+0.2}_{-0.3}$                  & $0.0^{+0.2}_{-0.4}$  & $0.0^{+0.2}_{-0.3}$ \\
   $\log U$               & ionisation parameter of the star-forming gas & $\mathcal{U}(-4, -1)$                                     & $-2.0^{+0.8}_{-0.8}$                 & $-2.3^{+0.8}_{-0.7}$ & $-2.4^{+0.8}_{-0.6}$ \\
   \midrule
   $\log SFR_{10} [\Msun \, \peryr]$ & star-formation rate averaged over the last 10~Myr & --- & $1.4^{+0.2}_{-0.4}$ & $1.3^{+0.2}_{-0.3}$ & $1.3^{+0.2}_{-0.4}$ \\
   $\log SFR_{100} [\Msun \, \peryr]$ & star-formation rate averaged over the last 100~Myr & --- & $1.2^{+0.4}_{-0.5}$ & $1.2^{+0.2}_{-0.3}$ & $1.1^{+0.4}_{-0.3}$ \\
   $age$ [Gyr]           & mass-weighted stellar age & --- & $0.23^{+0.7}_{-0.6}$ & $0.40^{+0.09}_{-0.08}$ & $0.38^{+0.10}_{-0.10}$ \\
   $EW(\OII)$ [\AA] & equivalent width of \OIIall & --- & $-30^{+20}_{-40}$ & $-40^{+20}_{-60}$ & $-40^{+25}_{-60}$ \\
   $EW(\Hbeta\!+\!\OIII)$ [\AA] & equivalent width of \Hbeta and \OIIIall & --- & $-130^{+80}_{-90}$ & $-140^{+60}_{-70}$ & $-130^{+70}_{-90}$ \\
   $EW(\Halpha\!+\!\NII)$ [\AA] & equivalent width of \Halpha and \NIIall & --- & $-310^{+170}_{-170}$ & $-330^{+130}_{-140}$ & $-320^{+150}_{-190}$ \\
   $D(Balmer)$ & Balmer-break index of \citet{binggeli+2019} & --- & $2.24^{+0.16}_{-0.18}$ & $2.25^{+0.15}_{-0.13}$ & $2.27^{+0.15}_{-0.20}$ \\
   $D_n 4000$  & 4000-\AA break index of \citet{balogh+1999} & --- & $1.14^{+0.03}_{-0.03}$ & $1.15^{+0.03}_{-0.02}$ & $1.17^{+0.02}_{-0.03}$ \\
   \midrule
   $z_\mathrm{obs}$       & redshift (alternative run) & $\mathcal{N}(z_\mathrm{phot}, 0.5)^b$   & $5.8^{+0.2}_{-0.1}$  & $5.9^{+0.1}_{-0.1}$ & $5.9^{+0.1}_{-0.1}$ \\
  \bottomrule
  \end{tabular}
  \end{center}

\raggedright{
(1) Parameter name (and units where applicable). (2) Parameter description. (3) Parameter prior probability distribution; $\mathcal{N}(\mu, \sigma)$ is the normal distribution with mean $\mu$ and dispersion $\sigma$; $\mathcal{U}(a, b)$ is the uniform distribution between $a$ and $b$; $\mathcal{T}(\mu, \sigma, \nu)$ is the Student's $t$ distribution with mean $\mu$, dispersion $\sigma$ and $\nu$ degrees of freedom; $\mathcal{G}(\mu, \sigma, a, b)$ is the normal distribution with mean $\mu$ and dispersion $\sigma$, truncated between $a$ and $b$.
(4) Median and 16\textsuperscript{th}--84\textsuperscript{th} percentile range of the marginalised posterior distribution for the model with the fiducial SFH prior (rising SFH); the nine log $SFR$ ratios are considered nuisance parameters so we do not report their posterior statistics. (5) Same as (4), but for the model with the `continuity' SFH prior. (6) Same
as (4), but for the model with the `bursty' SFH prior.
$^a$ The fiducial model uses a rising SFH prior \citep{turner+2025}; for each time bin $i$ in the SFH, the mean of the log SFR ratio $\Xi_i$ is set by the accretion rate of dark matter haloes \citep[][their eq.~4]{turner+2025}. For the continuity and bursty SFH priors, $\Xi_i=0$; for the continuity prior only, the scale parameter of the Student's $t$ prior is set to 0.3.
$^b$ This row shows the posterior redshift probability from alternative runs of \prospector where the only difference is the prior probability on the redshift; in this run, we assumed as prior $z_\mathrm{phot}$. All the resulting posterior probabilities are statistically consistent with the spectroscopic redshift.
}
\end{table*}

\subsection{A Low Mass Associated Satellite?}\label{s.sed.ss.satellite}

In the mass maps traced by F410M (rest-frame 6,000~\AA; Fig.~\ref{f.img.a}), the only other source in the vicinity of \target is 
the faint galaxy \satellite, located 1~arcsec (5.89 kpc) to the South of the main galaxy. The photometric redshift of 
this source is uncertain (the \eazy fit to the source is $z_{\mathrm{phot}} = 4.8^{+0.6}_{-0.4}$) owing to the faint photometry of the source in the \hst/ACS and NIRCam filters shortward of 2.0~\mum (Fig.~\ref{f.sed.g}). We do observe a clear drop between the F277W and F200W filters which is consistent with a strong Balmer break at the redshift $z=5.89$ of \target. Newer observations in F200W also reveal a
clear detection in F200W (9~\textsigma), which rules out the F277W break being a \Lyalpha break\footnote{For consistency with the rest of the article, the SED fitting uses the v0.9.5 photometry, where F200W is not detected. Updating to the deeper photometry gives a
younger SFH, but confirms a strong Balmer break.}. Interpreting the photometric drop 
as a Balmer break would be in agreement with the relatively flat F335M to F356W and F444W to F410M colors, which imply a lack of \OIIIall and \Halpha emission at this redshift.
Ironically, we cannot confirm if this galaxy is a faithful satellite of
\target or not, given the \eazy preference for a lower redshift which is primarily driven by the observed F090W - F115W color for the source.
Still, the spatial proximity and 
possible Balmer break compel us to explore the scenario of a post-starburst 
satellite. For this purpose, we extract the 
photometry of \satellite within a circular aperture of radius 0.25 arcsec (CIRC3 in the photometric 
catalogue, which we represent with a small blue circle in Fig.~\ref{f.sed.a}) and run the \prospector model inference with the same 
setup as \target. The resulting SED is shown in Fig.~\ref{f.sed.g} and the SFH is
in panel~\subref{f.sed.i}. Using the same prior probability distribution on the redshift as for \target, 
the inferred posterior distribution is $z = 5.3\pm0.3$, lower but still statistically consistent with the 
redshift of \target as well as the \eazy photometric redshift. Emboldened by this agreement, we repeat the inference using as redshift prior for \satellite 
the same spectroscopic prior for \target. The posterior probabilities of the parameters of this fiducial model
are reported in Table~\ref{t.others}. The mass is $\mstar = 10^{9.3}~\Msun$, six times smaller than 
\target.
The current and recent SFRs are both undetected, and the general shape of the SFH implies that this
galaxy, if confirmed to be at $z=5.89$, reached the peak of its SFH at $z=7\text{--}10$ (roughly 100~Myr before
\target) and then quenched permanently, reaching a sufficiently low SFR for 100~Myr to be classified as 
quiescent. Owing to the faint nature and lack of emission lines in \satellite, this scenario requires deep spectroscopy to be confirmed.

\begin{table}
    \begin{center}
    \caption{Summary of the posterior probabilities for the
    \prospector model parameters of the candidate quiescent satellite \satellite and for the East and West clouds.}\label{t.others}
    \setlength{\tabcolsep}{4pt}
    \begin{tabular}{lccc}
  \toprule
   Parameter$\,^a$                      & \satellite             & East cloud$^b$            & West cloud$^b$            \\
   \midrule                           
   $\log \mstar [\Msun]$              & $9.3^{+0.1}_{-0.1}$    & $8.39^{+0.06}_{-0.04}$    & $8.0^{0.3}_{-0.1}$        \\
   $\log Z [\Zsun]$                   & $-1.6^{+0.5}_{-0.3}$   & $-1.8^{+0.1}_{-0.1}$      & $-1.75^{+0.04}_{-0.05}$   \\
   $n$                                & $-0.7^{+0.4}_{-0.2}$   & $-0.98^{+0.06}_{-0.02}$   & $-0.9^{+0.7}_{-0.1}$      \\
   $\tau_V$                           & $0.5^{+0.2}_{-0.2}$    & $0.14^{+0.01}_{-0.02}$    & $0.30^{+0.13}_{-0.06}$    \\
   $\mu$                              & $1.0^{+0.3}_{-0.3}$    & $0.2^{+0.6}_{-0.2}$       & $0.4^{+0.4}_{-0.3}$       \\
   $\log Z_\mathrm{gas} [\Zsun]$      & $-0.9^{+0.9}_{-0.8}$   & $-0.70^{+0.01}_{-0.01}$   & $-0.69^{+0.02}_{-0.01}$   \\
   $\log U$                           & $-2.8^{+1.1}_{-0.8}$   & $-1.2^{+0.1}_{-0.1}$      & $-1.04^{+0.03}_{-0.07}$   \\
   \midrule                                 
   $\log SFR_{10} [\Msun \, \peryr]$  & $-1.0^{+0.8}_{-1.6}$   & $1.21^{+0.02}_{-0.01}$    & $0.73^{+0.12}_{-0.03}$    \\
   $\log SFR_{100} [\Msun \, \peryr]$ & $-0.8^{+1.3}_{-1.5}$   & $0.35^{+0.05}_{-0.03}$    & $-0.20^{+0.19}_{-0.05}$   \\
   $age$ [Gyr]                        & $0.30^{+0.19}_{-0.16}$ & $0.04^{+0.01}_{-0.01}$    & $0.05^{+0.02}_{-0.01}$    \\
   $EW(\OII)$ [\AA]                   & $-1^{+1}_{-14}$        & $-230^{+15}_{-18}$        & $-210^{+17}_{-8}$         \\
   $EW(\Hbeta\!+\!\OIII)$ [\AA]       & $-8^{+7}_{-32}$        & $-7200^{+200}_{-180}$     & $-7000^{+900}_{-450}$     \\
   $EW(\Halpha\!+\!\NII)$ [\AA]       & $-30^{+30}_{-150}$     & $-6100^{+180}_{-170}$     & $-5800^{+670}_{-350}$     \\
   $D(Balmer)$                        & $2.7^{+0.4}_{-0.5}$    & $1.06^{+0.01}_{-0.01}$    & $1.18^{+0.03}_{-0.03}$    \\
   $D_\mathrm{n} 4000$                         & $1.16^{+0.04}_{-0.04}$ & $0.971^{+0.001}_{-0.001}$ & $0.992^{+0.005}_{-0.004}$ \\
  \bottomrule
  \end{tabular}
  \end{center}

\raggedright{
$^a$ A description of the parameters and their prior probabilities (where relevant) is in Table~\ref{t.prosp}.
$^b$ We do not interpret the East and West clouds as star-forming galaxies; the parameters of the \prospector model are reported as a benchmark, and to illustrate that interpreting their emission as a galaxy SED results in implausible parameter combinations. In particular, the extremely small uncertainties on certain parameters (e.g., dust index $n$, $\log U$, mass-weighted age, D$_\mathrm{n}$4000) are due to the models being a poor match to the observations.
}
\end{table}

\subsection{Emission-line regions}\label{s.sed.ss.clouds}

The large emission-line regions surrounding \target are unlikely to be 
dominated by star-formation photoionization, as we will demonstrate in this Section.
Galaxies have not been observed to display
the combination of very large size and very high \OIIIall and \Halpha EWs
seen in the SED of these clouds (Fig.~\ref{f.sed.e} and~\subref{f.sed.j}).
Large emission-line region sizes are associated with high \mstar \citep{shen+2003,
vanderwel+2014,ormerod+2023,ito+2024,ji+2024c,martorano+2024,miller+2024,ward+2024}, while high EWs are
associated with young, low-mass galaxies.

To quantify the plausibility of star-formation photoionization,
we extract SEDs using large apertures encompassing most of the East and 
West emission-line regions (yellow and purple lines in Fig.~\ref{f.sed.a}).
The resulting models (Table~\ref{t.others}) favour low stellar masses ($10^{8.4}$ and $10^{8}~\Msun$
respectively for the East and West clouds) with extremely high SFR
of 17 and 5~\Msun~\peryr. Given their inferred stellar masses and SFR values, these clouds would lie 1.5 and 1.7~dex above the star-forming main sequence (SFMS) at $z=5.89$
\citep{popesso+2023}.
The inferred dust attenuation is also very implausible.
The diffuse attenuation (affecting both birth clouds and 
older stars) is moderate, with $\tau_V = 0.15$ and 0.27, respectively. However, the ratio $\mu$
between birth-cloud and foreground attenuation is only 0.3, while typical values are around $\mu = 1$ \citep{charlot+fall2000,calzetti+2000}. The low value of $\mu$ arises from the competing requirements of achieving high continuum attenuation, necessary to suppress weak/undetected UV light, while maintaining low emission-line attenuation to preserve the brightness of the emission lines.
Similarly, the dust curve is as steep as possible (given the top-hat prior probability cut at $n\geq-1$), because with 
$n=-1$, the model again attempts to minimize the UV continuum without suppressing the rest-frame optical emission lines.

Nevertheless, even this contrived model cannot fully reproduce the data.
For both clouds, the UV continuum is over-predicted by up to 
7~\textsigma (Figs.~\ref{f.sed.f} and~\subref{f.sed.k}).
Moreover, the model clearly over-predicts the optical
continuum traced by F410M (by about 10~\textsigma in both cases).
Our parametric SFH approach with Student's $t$ probability 
priors leaves enough flexibility to reproduce sharp
bin-to-bin variations in the SFR, because the fat tails on
the prior probability are certainly less constraining than
the poorly fit 7--10 \textsigma Gaussians discussed above.
Therefore, these major model mismatches suggest that the star-forming interpretation is either highly incomplete or outright incorrect, with the nebular emission
driven by entirely different energy sources.

\section{AGN modelling}\label{s.agn}

\begin{figure}
  {\phantomsubcaption\label{f.cigale.a}
   \phantomsubcaption\label{f.cigale.b} 
   \phantomsubcaption\label{f.cigale.c} 
   \phantomsubcaption\label{f.cigale.d}}
  \includegraphics[width=\columnwidth]{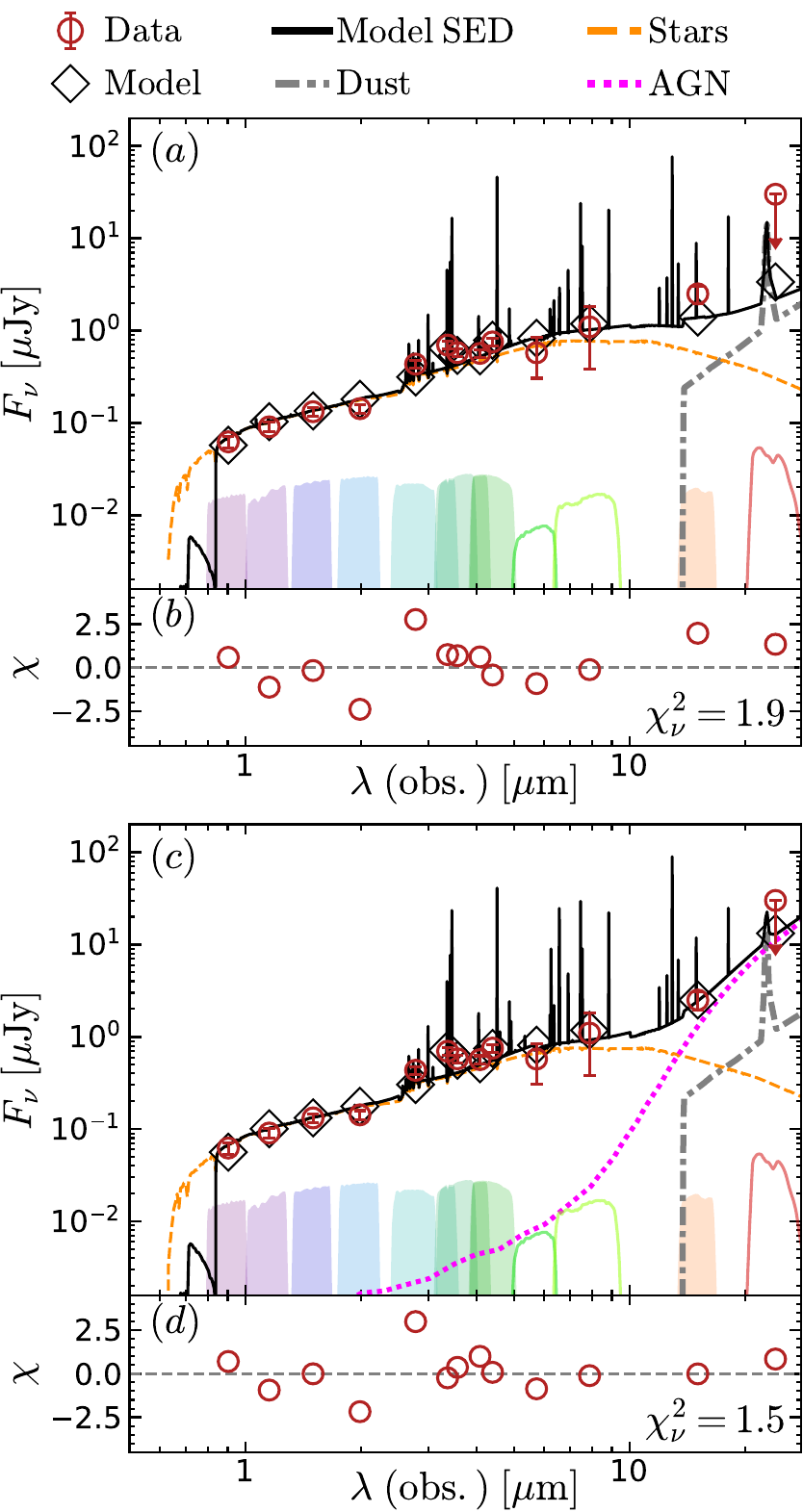}
  \caption{
  \cigale SED modelling of the main galaxy \target, contrasting the model without AGN 
  (panels~\subref{f.cigale.a} and~\subref{f.cigale.b}) and with AGN 
  (panels~\subref{f.cigale.c} and~\subref{f.cigale.d}).
  The filled filter transmission curves are \jwst/NIRCam
  and MIRI filters, while the empty transmission curves
  are \textit{Spitzer}/IRAC and MIPS.
  The two models are statistically consistent, as
  inferred from comparing the reduced $\chi^2_\nu$, yet the AGN model is better
  at reproducing the MIRI F1500W flux, which \cigale
  interprets as evidence for an obscured AGN. The MIR
  data places an upper limit on the AGN luminosity of
  $L_\mathrm{bol} \lesssim 10^{45.8} \, \mathrm{erg\;s^{-1}}$,
  }\label{f.cigale}
\end{figure}

To test for on obscured AGN, we use the SED fitting tool \cigale \citep{yang+2022}, fitting to the Kron aperture NIRCam photometry to the source. At longer wavelengths, we include \textit{ad-hoc}
photometric measurements from both \textit{Spitzer}/IRAC ch3 and ch4 and
\jwst/MIRI (Section~\ref{s.obs}). These fluxes are important as they help to constrain the possibility of an obscured AGN that may only be visible in the near-to-mid IR. For \cigale, we employ the \textsc{skirtor} AGN implementation from \citet{stalevski+2016}, while the host galaxy stellar-population templates are
from \citet{bruzual+charlot2003}. We used the default
SFH, parametrized by a delayed exponential.
Dust attenuation assumes the \citet{gordon+2003}
law for the AGN, and the \citet{calzetti+2000} law for the stellar populations.
The resulting fits are shown in Fig.~\ref{f.cigale}, where we contrast the
model where both include and don't include emission from an AGN (panels~\subref{f.cigale.a} and~\subref{f.cigale.c},
respectively). Both models are consistent with the observations within the uncertainties, with
reduced $\chi^2_\nu$ of 1.9 and 1.5 for the model without and with an AGN. However, the AGN model better reproduces the \jwst/MIRI flux at F1500W. Overall, the MIR observations are useful to constrain the luminosity 
of an obscured AGN. The AGN model returns a dust
attenuation towards the AGN of $A_V = 4.5\pm2.9$~mag. The luminosity is $\log (L_\mathrm{bol} \, [\mathrm{erg\;s^{-1}}]) = 45.8\pm0.6$~dex, which we treat as an upper limit 
given the non detection by \textit{Spitzer}/MIPS at 24~\mum and the low reduced $\chi^2_\nu$ of the non-AGN model.
There is a large discrepancy between the data and model around the Balmer-break region,
where both \cigale model fits under-predict the strength of the Balmer break. This is likely
due to our SFH setup in \cigale, which lacks the flexibility for allowing a period of
low SFR followed by a `crescendo', as we discussed in Section~\ref{s.sed.ss.central}. Overall, this discrepancy results in a strong disagreement about the
recent SFH, with \cigale inferring that the SFR averaged
over the 10~Myr prior to observation is $260\pm15~\Msun\,yr^{-1}$. In contrast, the stellar mass is in good agreement with \prospector, with \cigale finding $\log(\mstar/\Msun) = 10.06\pm0.02$, with minimal differences if an AGN is included or not.

\section{Emission-line luminosity}\label{s.eml}

From the surface brightness distribution of the emission-line regions, we can 
estimate the flux and equivalent width of the main emission-line groups,
\Hbeta and \OIIIall in F335M and F356W, and \Halpha in F410M and 
F444W. For the rest-frame equivalent width of \Hbeta and \OIIIall, since the lines fall in both the medium- and wide-band filters, we use the expression
\begin{align}\label{eq.ewo3}
   EW(\Hbeta + \OIII) & \approx \dfrac{\rekt{F335M}}{1+z} \dfrac{1 - \dfrac{1}{\delta_\mathrm{cont}} \dfrac{\lambda_{F356W}^2}{\lambda_{F335M}^2} \dfrac{F_{\nu\,F335M}}{F_{\nu\,F356W}}}{
       1 - \dfrac{\lambda_{F356W}^2}{\lambda_{F335M}^2} \dfrac{\rekt{F335M}}{\rekt{F356W}} \dfrac{F_{\nu\,F335M}}{F_{\nu\,F356W}}}\\
   \delta_\mathrm{cont} & \equiv \dfrac{\lambda_{F356W}^2}{\lambda_{F335M}^2} \dfrac{F_{\nu\,F335M,\, {\rm cont}}}{F_{\nu\,F356W,\,{\rm cont}}},
\end{align}
where $\lambda_{F335M}$ is the pivot wavelength of the F335M filter, \rekt{F335M} is the rectangular width of the 
F335M filter (defined as the effective width of the filter divided by its maximum throughput), and
$\delta_\mathrm{cont}$ is a continuum-colour correction term, whose value we do not know, due to the lack of
sufficient medium bands to characterize the continuum shape (see Appendix~\ref{s.linephot}
 for a derivation). For reference, for a
flat continuum in $F_\lambda$, $\delta_\mathrm{cont} \approx 1$. The resulting equivalent width has the same
units as \rekt{F335M}. For \Halpha, which falls in the wide-band filter but not in
the medium-band filter, we have
\begin{align}\label{eq.ewha}
   EW(\Halpha) & \approx \dfrac{\rekt{F444W}}{1+z} \left(1 - \delta_\mathrm{cont} \cdot \dfrac{\lambda_{F410M}^2}{\lambda_{F444W}^2} \dfrac{F_{\nu\,F444W}}{F_{\nu\,F410M}} \right)\\
   \delta_\mathrm{cont} & \equiv \dfrac{\lambda_{F444W}^2}{\lambda_{F410M}^2} \dfrac{F_{\nu\,F410M,\, {\rm cont}}}{F_{\nu\,F444W,\,{\rm cont}}},
\end{align}
These approximations are somewhat simplistic: because the continuum spectral density 
features in the denominator of the EW definition, the errors of Eqs.~\ref{eq.ewo3} and~\ref{eq.ewha} can be large for any range of EW values. These errors are driven 
primarily by the colour of the continuum between the two photometric bands used.

In contrast, the corresponding emission-line fluxes are more stable for regions with higher EW values. For \Hbeta and \OIIIall, we have
\begin{equation}\label{eq.o3sb}
\begin{split}
   F(\Hbeta + \OIII) \approx 10^{-5} \dfrac{\mathrm{c}}{\lambda_{F335M}^2} & \dfrac{\rekt{F335M}\;\rekt{F356W}}{\rekt{F356W} - \rekt{F335M}} \\
   \times & \left(F_{\nu\;F335M} - \dfrac{\lambda_{F335M}^2}{\lambda_{F356W}^2} F_{\nu\;F356W} \right),
\end{split}
\end{equation}
while for \Halpha we have
\begin{equation}\label{eq.hasb}
\begin{split}
   F(\Halpha) \approx 10^{-5} \dfrac{\mathrm{c}}{\lambda_{F444W}^2} & \rekt{F444W} \\
   \times & \left(F_{\nu\;F444W} - \dfrac{\lambda_{F444W}^2}{\lambda_{F410W}^2} F_{\nu\;F410M} \right),
\end{split}
\end{equation}
where the pivot wavelengths and rectangular widths are in units of \mum, c is
the speed of light in \kms, $F_\nu$ are spectral flux densities in nJy, and the resulting 
emission-line flux is in units of \fluxcgs[-18][]. Identical expressions apply to
surface brightness, by replacing $F_\nu$ with ${SB}_\nu$.

\begin{figure}
  \includegraphics[width=\columnwidth]{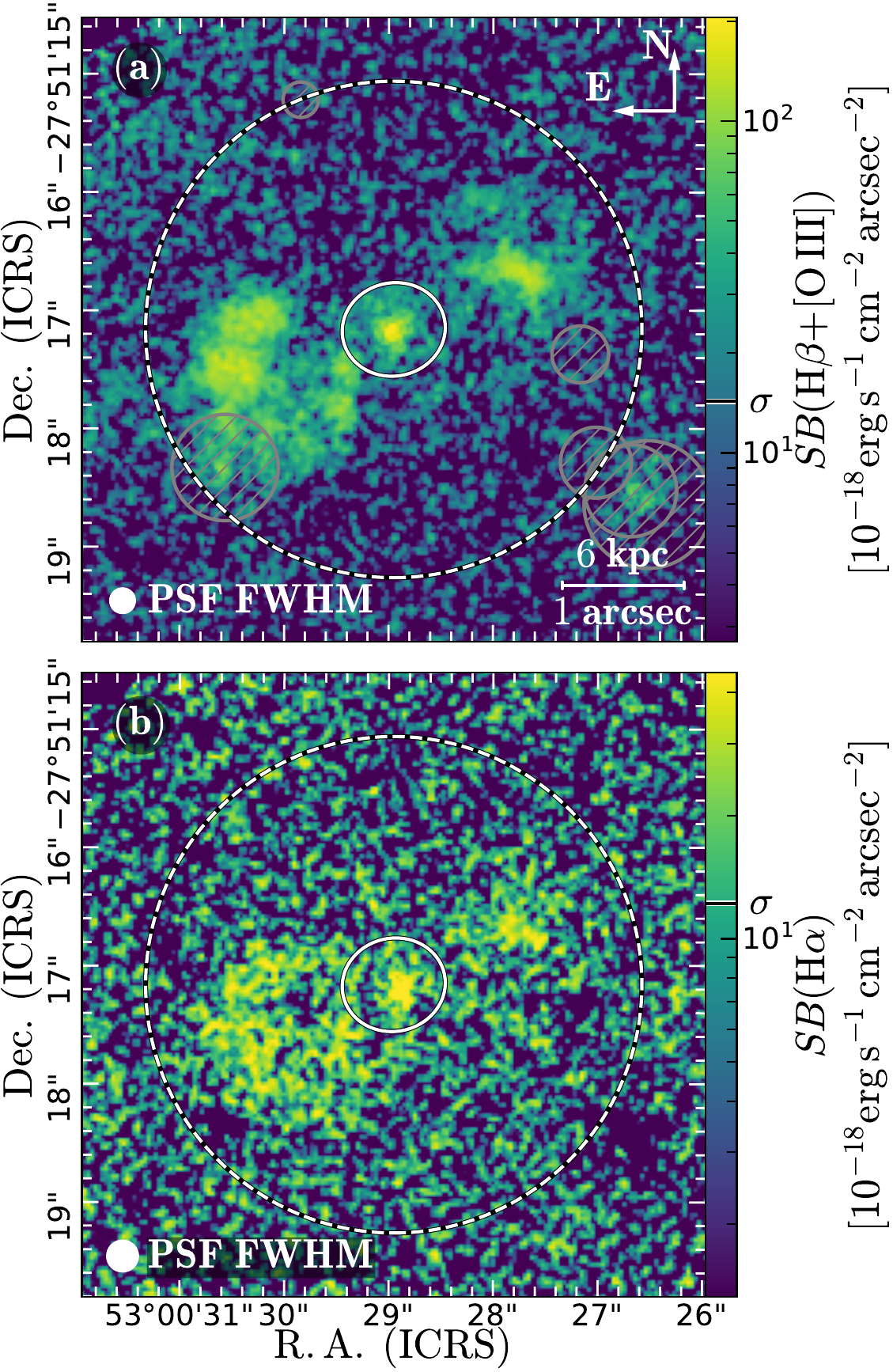}
  {\phantomsubcaption\label{f.emlmap.a}
   \phantomsubcaption\label{f.emlmap.b}}
  \caption{NIRCam-derived emission-line maps of \Hbeta and \OIIIall (panel~\subref{f.emlmap.a}), and \Halpha (panel~\subref{f.emlmap.b}).
  The colourbar shows the 60\textsuperscript{th}--99.9\textsuperscript{th}
  percentile range, with the 1-\textsigma noise level highlighted by a
  horizontal mark.
  The dashed circle is the region where we estimate the total flux; the solid
  ellipse is excluded and is considered to be associated with the galaxy (Section~\ref{s.morph}).
  The observed \OIII luminosity of the nebula exceeds that of the host galaxy by an order of
  magnitude. Note the different morphology of the host galaxy in the two images;
  differential concentration of \OIIIall vs \Halpha may indicate obscured AGN activity. The PSFs FWHM in
  F356W and F444W are indicated by the white circle. 
  }\label{f.emlmap}
\end{figure}

The surface brightness maps obtained by applying eq.s~\ref{eq.o3sb} and~\ref{eq.hasb} to 
the JADES data are displayed in Fig.~\ref{f.emlmap}. The \Hbeta+\OIII map 
(panel~\subref{f.emlmap.a}) and \Halpha map (panel~\subref{f.emlmap.b}) reach
surface-brightness noise levels of 1.4 and 1.2\sbcgs[-17], corresponding to
5-\textsigma flux sensitivities of 6 and 5\fluxcgs[-19] in a circular aperture with 0.15-arcsec radius.
The total area where flux is detected in \Hbeta+\OIII only is 44.6~kpc$^2$.

To measure total line luminosities, we define two apertures for estimating the emission from
the host galaxy and from the nebula. For the host, we define an elliptical aperture 
with the position angle and shape of the stars in the host galaxy (Section~\ref{s.morph}), reaching 2~\re (solid line in 
Fig.~\ref{f.emlmap}); for the nebula, we use a much larger circular aperture of radius 
2.1~arcsec (dashed line).
For the nebula, we measure fluxes of $F(\Hbeta+\OIII) = (163\pm2)$~\fluxcgs[-18] and $F(\Halpha) =
(60\pm1)$~\fluxcgs[-18]. For comparison,
integrating the \Halpha emission lines from the SAPPHIRES apertures and summing the two clouds we find 
$F(\Halpha) = (18\pm2)$~\fluxcgs[-18], implying a
flux mismatch of about three. This is not surprising, 
and may arise from a combination of inaccurate flux measurement from NIRCam photometry, plus low sensitivity
of NIRCam/WFSS to diffuse and broad line emission.
Neglecting dust attenuation, the luminosities of the host galaxy in each 
of the two emission-line groups \Hbeta+\OIII and \Halpha are $L_{\Hbeta+\OIII}
=(2.10\pm0.04)\times 10^9~\mathrm{L_\odot}$ and
$L_{\Halpha}=(0.86\pm0.04)\times 10^9~\mathrm{L_\odot}$. The 
nebula (without the host galaxy) has $L_{\Hbeta+\OIII}=(1.78\pm0.02)\times 10^{10}~\mathrm{L_\odot}$ and
$L_{\Halpha}=(0.66\pm0.02)\times 10^{10}~\mathrm{L_\odot}$.
In \Hbeta+\OIII, the 
nebula is 8 times more luminous than the host galaxy. Even in the extreme scenario where 
the host galaxy is corrected for significant dust attenuation ($\tau_V \cdot (1+\mu) 
\approx 2.7$; Table~\ref{t.prosp}\footnote{For a single dust screen, one has $A_V = 2.5 \log(e) \tau_V$, but since \prospector uses two dust screens, the exact conversion between $\tau_V$ and $A_V$ depends on the ratio between young and old stellar populations, where the former are attenuated by both the diffuse and `birth-cloud' dust, while the latter are attenuated only by diffuse dust.}) while assuming no dust for the nebula, the luminosity 
ratio would still be 0.5, making the nebula
comparable to the whole galaxy.
To place these measurements in context, we estimate the
\Hbeta luminosity from the Balmer decrement
$L_{\Halpha}/L_{\Hbeta}=2.86$, assuming Case~B recombination, $T_\mathrm{e}=10,000$~K and
electron density $n_\mathrm{e}=100~\mathrm{cm}^{-3}$ \cite{osterbrock+ferland2006}. With this assumption,
an \OIII doublet ratio of 2.98 and no dust attenuation,
we estimate $L_{\OIIIL} \sim 1.2\times10^{10}~\mathrm{L_\odot} = 4.6\times 
10^{43}$~erg~s$^{-1}$.
With this value, we derive $L_{\OIIIL}/L_{\Hbeta} \approx 5$, lower than the narrow-line regions of quasars at $z=0.4\text{--}0.7$ \citep[both obscured and unobscured;][]{liu+2013,liu+2014}.

Alternatively, if we assume the typical $\OIIIL/\Hbeta \sim 10$ of quasar-illuminated nebulae \citep{liu+2013,liu+2014,lyu+2024} or outflows \citep{scholtz+2020}, we would infer an \Hbeta luminosity of $1.2\times10^9~\mathrm{L_\odot}$, 2 times lower than the prediction from
the observed \Halpha luminosity.
The grism spectra suggests that the contribution of \NIIall to the nebular emission is negligible (Section~\ref{s.redshift}), which would imply some degree
of dust attenuation. However, a weak continuum could also
alter the line ratios, so confirming the presence of
dust requires deeper, spatially resolved spectroscopy.

Regardless of the estimate of $L_{\Hbeta}$, the resulting
\OIIIL luminosity is extremely high for a non-quasar galaxy; typical unobscured quasars at $z=0.4\text{--}0.7$ 
have $L_{\OIIIL} \approx 2.5\times10^9~\mathrm{L_\odot}$ \citep{liu+2014}, but \target
shows no trace of an active quasar at the centre of the host galaxy. \citet{hainline+2013} compared the \OIIIL and rest-frame 8-\mum 
luminosity of a sample of 50 obscured quasars; using their empirical relation (their 
fig.~5), we infer that \target could have a flux density as high as $S_{50~\mum} \sim 
10$~\textmu Jy, which is in the range of predictions
of our \cigale AGN models (Fig.~\ref{f.cigale}).

With $L_{\Halpha} = 0.66\times10^{10}~\mathrm{L_\odot}$, assuming no dust attenuation and all of the line emission arising from star formation, and adopting the \citet{kennicutt+evans2012} conversion, the nebula would have a $SFR \sim 80$~\Msun~\peryr, which is larger than the host galaxy
(Fig.~\ref{f.sed.c}).

\section{Large-scale environment}\label{s.environment}

Early massive galaxies, such as \target, are thought to trace some of the most active star-forming regions in the Universe. As a result of this, dusty star-forming galaxies (DSFGs) and sub-millimetre galaxies (SMGs) have often been used to identify galaxy protoclusters at high redshifts \citep{chapman+2009,riechers+2014,umehata+2015,oteo+2018,long+2020,sunf+24,fudamoto+2025}. One reason for this is the growing evidence for accelerated galaxy evolution within overdense environments 
\citep[e.g.,][]{helton+2024a,helton+2024b,arribas+2024,morishita+2024}.
DSFGs and SMGs are some of the most extreme objects found at high redshifts, so it would make sense for these to be situated within an overdense region of galaxies.

Indeed, there is already spectroscopic evidence for an over-density in GOODS-S around $z=5.8\text{--}5.9$, based on Keck redshifts from \Lyalpha emission for $i$-band drop galaxies in this redshift range \citep{bunker+2003,stanway+2004a,stanway+2004b}.

To explore the large-scale environment surrounding \target, we searched for relatively bright galaxies ($m < 28.5\ \mathrm{AB\ mag}$ in F444W, adopting Kron photometry) that are consistent with the redshift of \target by using the v0.9.5 JADES GOODS-S photometric catalogue \citep{rieke+2023}. We measured photometric redshifts with \eazy \citep{brammer+2008} following the methodology of \citet{hainline+2024} using CIRC1 photometry (0.10-arcsec radius) and adopting the redshift where the likelihood is maximized ($\chi^{2}$ is minimized), which produced $z_{\mathrm{phot}} = 6.01$ for \target. The median photometric redshift uncertainty at $z_{\mathrm{phot}} \approx 6$ is $\Delta z_{\mathrm{phot}} \approx 0.24$ (corresponding to velocity uncertainties of roughly $10,000\ \mathrm{km/s}$). Photometric redshift uncertainties are measured as the difference between the 16\textsuperscript{th} and 84\textsuperscript{th} percentiles of the photometric redshift posterior distributions. We choose to not make any cut on the photometric redshift uncertainty since this would bias against dusty and quiescent galaxies like \target and the candidate satellite galaxy. Based on the typical photometric redshift uncertainty, we select galaxies that are within $\Delta z_{\mathrm{phot}}$ of \target (i.e., $5.77 < z_{\mathrm{phot}} < 6.25$). There is some evidence for the uncertainties on $z_\mathrm{phot}$ from \eazy being under-predicted, but investigating this possibility goes beyond the scope of this article. Following these selections, we are left with a photometric sample of $N = 998$ sources at $z_{\mathrm{phot}} \approx 6$. We identify nine sources within 2 arcsec from \target, of which six were selected in the sample of 998; including these six sources in the clustering analysis would increase the significance of our results. However, since these sources are most likely gas clouds and not galaxies, we conservatively remove them from the sample.

Following the methodology outlined in \citet{sandles+2023} and \citet{alberts+2024}, we utilize a kernel density estimator (KDE) to estimate the underlying density field for the photometric sample. We assumed Gaussian kernels and optimize the assumed bandwidth (i.e., the smoothing scale) by maximizing the likelihood cross-validation quantity \citep{chartab+2020}. The optimized bandwidth for the photometric sample is $2.402\ \mathrm{cMpc}$ and corresponds to roughly $1.0\ \mathrm{arcmin}$ at the redshift of \target. We further implement a correction factor to compensate for KDEs underestimating densities near the edge of footprints \citep{taamoli+2024}. We finally subtract the individual density values for each galaxy by the mean density value estimated across the entire field to ensure that only half of the photometric sample is classified as overdense.

\subsection{A new candidate overdensity in GOODS-S}\label{s.environment.ss.overd}

The adopted methodology using the KDE identifies two spatially distinct galaxy overdensities that have peak significance levels higher than 3~\textsigma, where \textsigma is defined to be the standard deviation of the density values across the entire field. The former of these two overdensities is at R.A. 53.1297, Dec. $-27.7854$ with a peak density of $3.3$ times that of a random volume. This peak corresponds to the spectroscopically confirmed galaxy overdensity JADES-GS-OD-5.928 \citep{helton+2024b}. The second of these overdensities is at R.A.~53.0239, Dec.~$-27.84233$
with a peak density of $3.8$ times that of a random volume. It consists of 44 galaxies (including \target), with mean redshift $z_\mathrm{phot}=5.99$ and a standard deviation of 0.04.
The centre of this new overdensity is sufficiently far from previously known \Lyalpha emitters \citep{stanway+2004b} to be considered separate.
If we include the six sources identified as clouds in the analysis, the significance would increase to 4.1 times the random value, and the number of galaxies would increase from 44 to 56 (since the size of overdensity also increases slighly).
Our inferred centre is separated by only $2.760\ \mathrm{cMpc}$ ($1.126\ \mathrm{arcmin}$) from \target (at the centre of the overdensity we find a close group of five galaxies; Appendix~\ref{s.centre}). However, since \target is near the edge of the footprint for which we have imaging from \hst and \jwst, it is possible that there is a substantial number of galaxies missed by our catalogue and therefore by the KDE. These missing galaxies would bias the KDE estimate of the peak density away from \target and towards the North, as observed.
Nevertheless, while this overdensity is photometrically even more significant than JADES-GS-OD-5.928, it lacks spectroscopic confirmation and is thus considered a candidate overdensity.

\begin{figure*}
  \includegraphics[width=\textwidth]{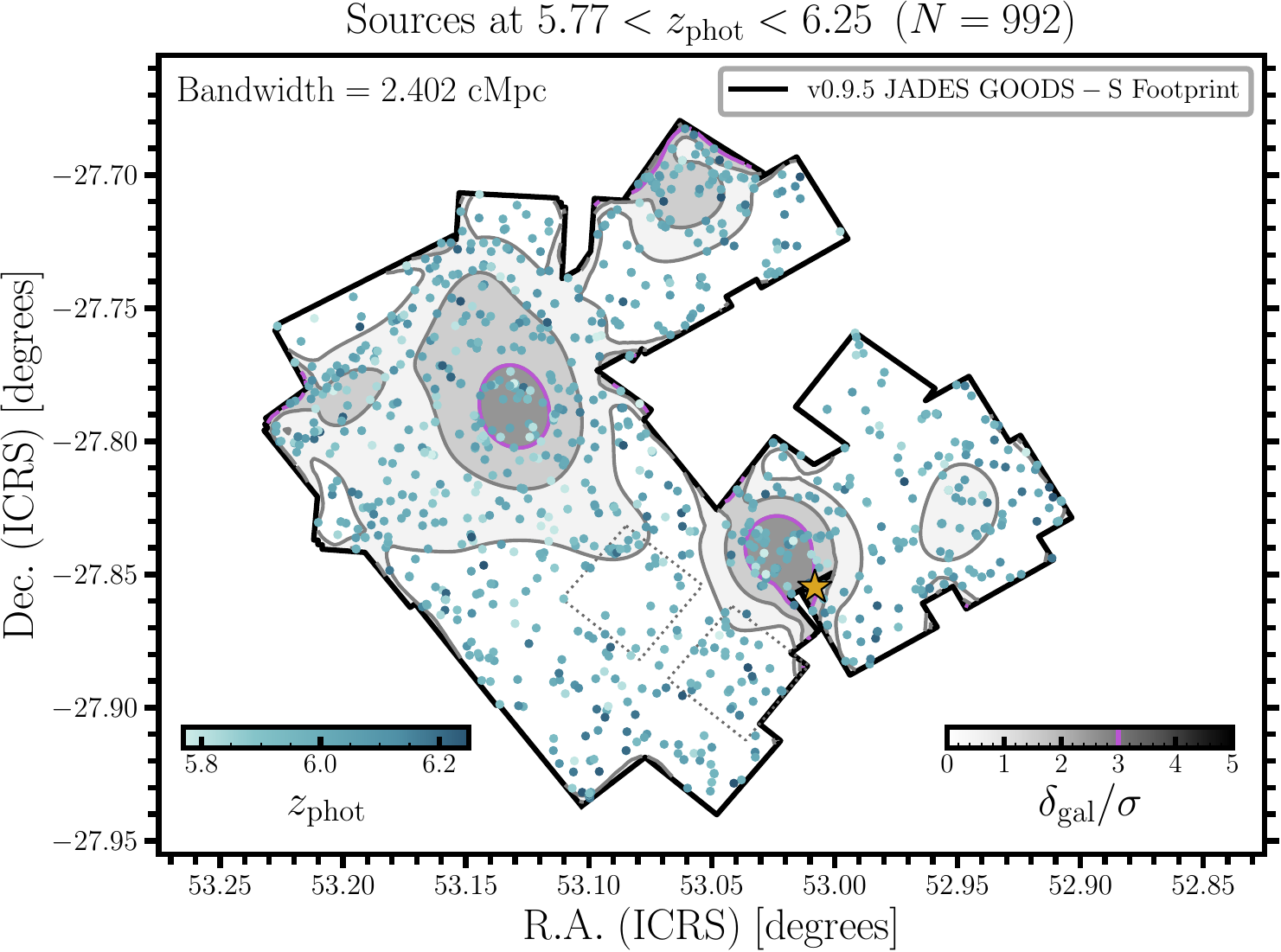}
  \caption{The spatial distribution of the $N = 992$ galaxy candidates in the photometric sample using points that are colour-coded by their photometric redshifts. These galaxy candidates are at $5.77 < z_{\mathrm{phot}} < 6.25$  in GOODS-S and are selected based on the available \hst and JWST photometry. The thick black polygon is the GOODS-S NIRCam footprint of JADES, while the dotted grey contours trace the JADES Origins Field, the deepest NIRCam imaging. The underlying density field is estimated with a KDE and is illustrated by the contours. The contours increment by 1~\textsigma, where \textsigma is the standard deviation of the density values across the entire field. The purple contour represents a significance level of 3~\textsigma, which is the threshold we adopt to identify galaxy overdensities. We identify two spatially distinct galaxy overdensities. One of these corresponds to a spectroscopically confirmed galaxy overdensity in GOODS-S (JADES$-$GS$-$OD$-5.928$ from \citealp{helton+2024b}). The other one of these is only $2.311\ \mathrm{cMpc}$ ($0.943\ \mathrm{arcmin}$) in separation from \target. The gold star illustrates the location of \target, which falls near the edge of the JADES footprint, shown by the black line.}\label{f.overd}
\end{figure*}

Fig.~\ref{f.overd} illustrates the spatial distribution of the $N = 992$ galaxy candidates in the photometric sample using points that are colour-coded by their photometric redshifts. These galaxy candidates are at $5.77 < z_{\mathrm{phot}} < 6.25$  in GOODS-S and are selected based on the available \hst and \jwst photometry. The underlying density field is estimated with a KDE and is illustrated by the contours. The contours increment by 1~\textsigma, where \textsigma corresponds to the standard deviation of the density values across the entire field. The purple contour represents a significance level of 3~\textsigma, which is the threshold we adopt to identify galaxy overdensities. The gold star illustrates the location of \target, which falls near the edge of the JADES footprint, shown by the thick black line.

Given the proximity with one of the extreme galaxy overdensities identified here, it is possible that the large-scale environment is in some part linked to the extended line-emitting nebulae surrounding \target. Environmental effects may have caused accelerated galaxy evolution and the creation of an AGN within \target that is responsible for the expelled gas, or they could have increased the rate of mergers and interactions that has tidally stripped the gas from \target. Either way, it is likely that environment has played an important role, and it is possible that \target is a brightest cluster galaxy (BCG) in the making. Future spectroscopic observations will provide more information about the large-scale structure surrounding \target and the impact of environment.

\subsection{Environment impact on galaxy properties}\label{s.environment.ss.effects}

\begin{figure}
  \includegraphics[width=\columnwidth]{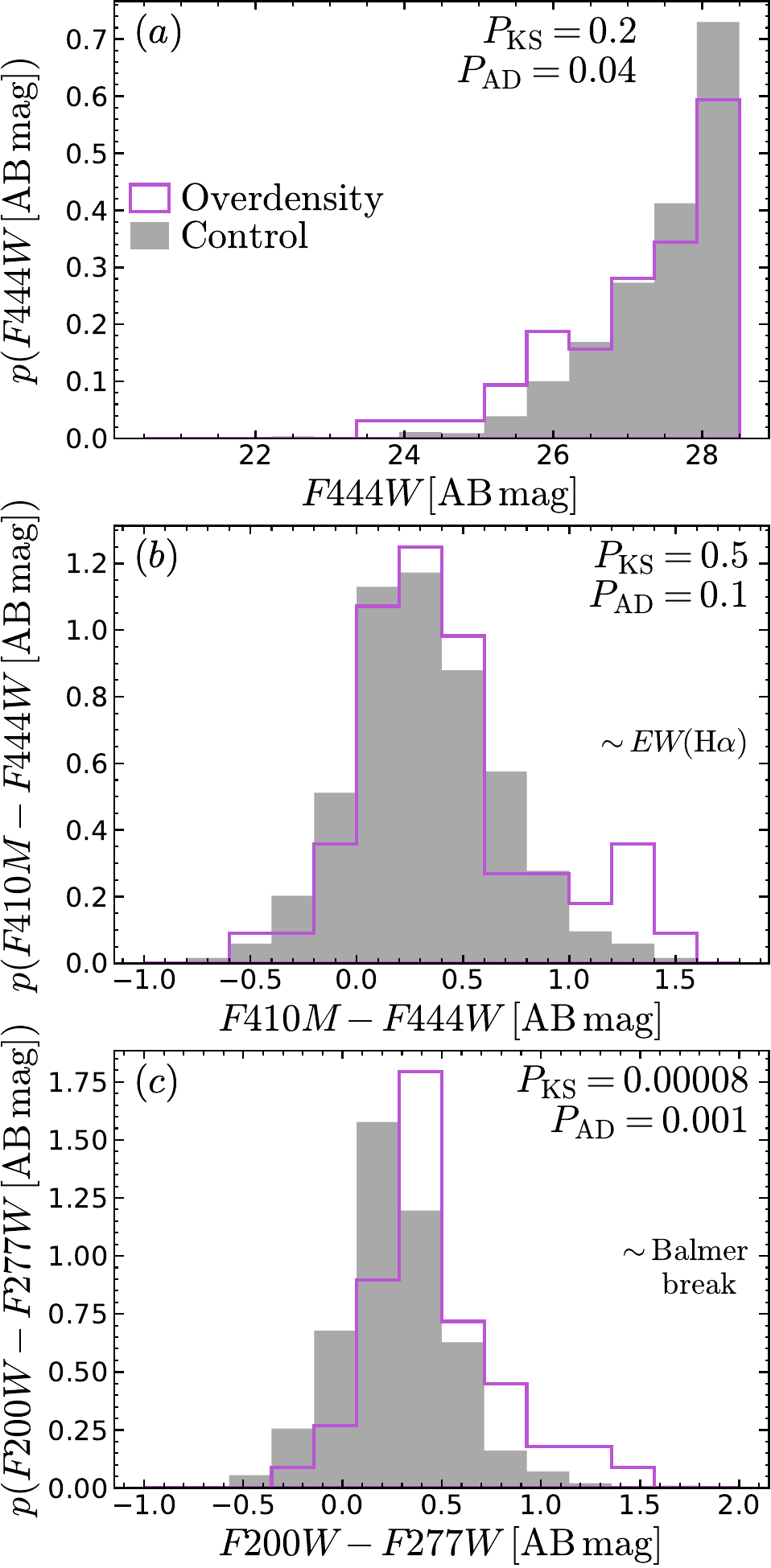}
  {\phantomsubcaption\label{f.env.emp.a}
   \phantomsubcaption\label{f.env.emp.b}
   \phantomsubcaption\label{f.env.emp.c}}
  \caption{Comparing the properties of 56 galaxies in the candidate overdensity (purple) to the control sample of 914 galaxies not in any overdensity (filled grey). We report the p-value $P$ of the null hypothesis that the two samples are drawn from the same distribution (according to the KS and AD tests). We find no difference in F444W magnitude and
  in F410M-F444W colour, but a statistically significant excess in F200W-F277W colour (tracing the Balmer break at $z\sim6$), likely
  indicating that galaxies in overdensities have on average older 
  light-weighted ages.
  }\label{f.env.emp}
\end{figure}

To gauge the possible effects of large-scale environment on the SFH of \target, in Fig.~\ref{f.env.emp} we compare the properties of the 56 galaxies in the candidate overdensity (purple empty histogram) to the control sample of 914 galaxies that are not in any known overdensity (filled grey histogram)\footnote{For a  test of small-scale environment effects see Appendix~\ref{s.centre}.}.
To test if the two samples are consistent with being drawn from the same distribution, we use a Kolmogorov-Smirnov (KS) test (but we
also report the p-value from Anderson-Darling, AD test).
Fig.~\ref{f.env.emp.a} compares the distributions of F444W magnitude, which at $z\sim6$ traces the rest-frame red optical, including \Halpha emission. We find no statistical evidence of
a difference between the two distributions.
Similarly, we find no evidence of a difference in the F410M-F444W 
colour (Fig.~\ref{f.env.emp.b}), which at $z\sim6$ traces primarily 
the EW of \Halpha (Eq.~\ref{eq.ewha}; \NIIall and \SIIall are much fainter than \Halpha at $z>5$, \citealp{cameron+2023b,sandles+2024}). In contrast, we find strong evidence of a colour difference in F200W-F277W, the two NIRCam bands bracketing the Balmer break at $z\sim 6$ (Fig.~\ref{f.env.emp.c}; p-value $P<10^{-4}$). This remarkable difference in
F200W-F277W colour suggests that galaxies in overdensities have on average older light-weighted stellar-population ages than galaxies in non-overdense regions, consistent with an earlier start of their
SFHs.
Stellar metallicity could also play some role, due to the physical degeneracy between stellar age and metallicity, but a significant 
metallicity enhancement on large scales seems less likely at these
early epochs \citep[but has been observed in clusters of galaxies at lower redshifts $z=0.35$,][]{gupta+2016}. A contribution from emission lines (chiefly \OIIall and \NeIIIL) is also possible, but it is unlikely to be the dominant effect. First, the large values of F200W-F277W (reaching 1~mag; Fig.~\ref{f.env.emp.c}) strongly suggest that a difference in continuum shape must be present, at least in some galaxies (see also discussion about the Balmer break in \target; Section~\ref{s.sed.ss.central}).
Second, we detect no enhancement in F410M-F444W (Fig.~\ref{f.env.emp.b}), so the EW of emission lines between galaxies in overdensities and the control must be similar.
Therefore, increasing F200W-F277W at fixed $EW(\Halpha)$ requires
very low ionization and/or high metallicity, which seems unlikely at $z\sim6$ and on large scales.
If the F200W-F277W is pointing to stronger Balmer breaks, then this
would imply a higher average mass-to-light ratio for galaxies in
overdensities, meaning that the similar distribution in F444W magnitude may hide different distributions in \mstar, that is,
at fixed luminosity, galaxies in the overdensity are more massive 
than those in the control sample.

To better characterize the physical properties of the galaxies in the candidate overdensity, we use \prospectorbeta \citep{wang+2023c}, a version of \prospector \citep{johnson+2021} optimized for redshift recovery. \prospectorbeta utilizes a series of galaxy evolution priors, including a stellar mass prior derived in \citet{leja+2020}, a prior on galaxy number density as a function of redshift, and the continuity prior for the star-formation history described in Section \ref{s.sed}, but here with 7 logarithmically-spaced time bins. In addition, there is a mass-dependence on the SFH prior, where the start of each age bin is shifted based on the stellar mass, as described in \citet{wang+2023c}. We refer the reader here for a complete description of the \prospectorbeta priors.
For consistency with the clustering analysis, the redshift is fixed to the photometric
redshift inferred with \eazy.

\begin{figure}
  \includegraphics[width=\columnwidth]{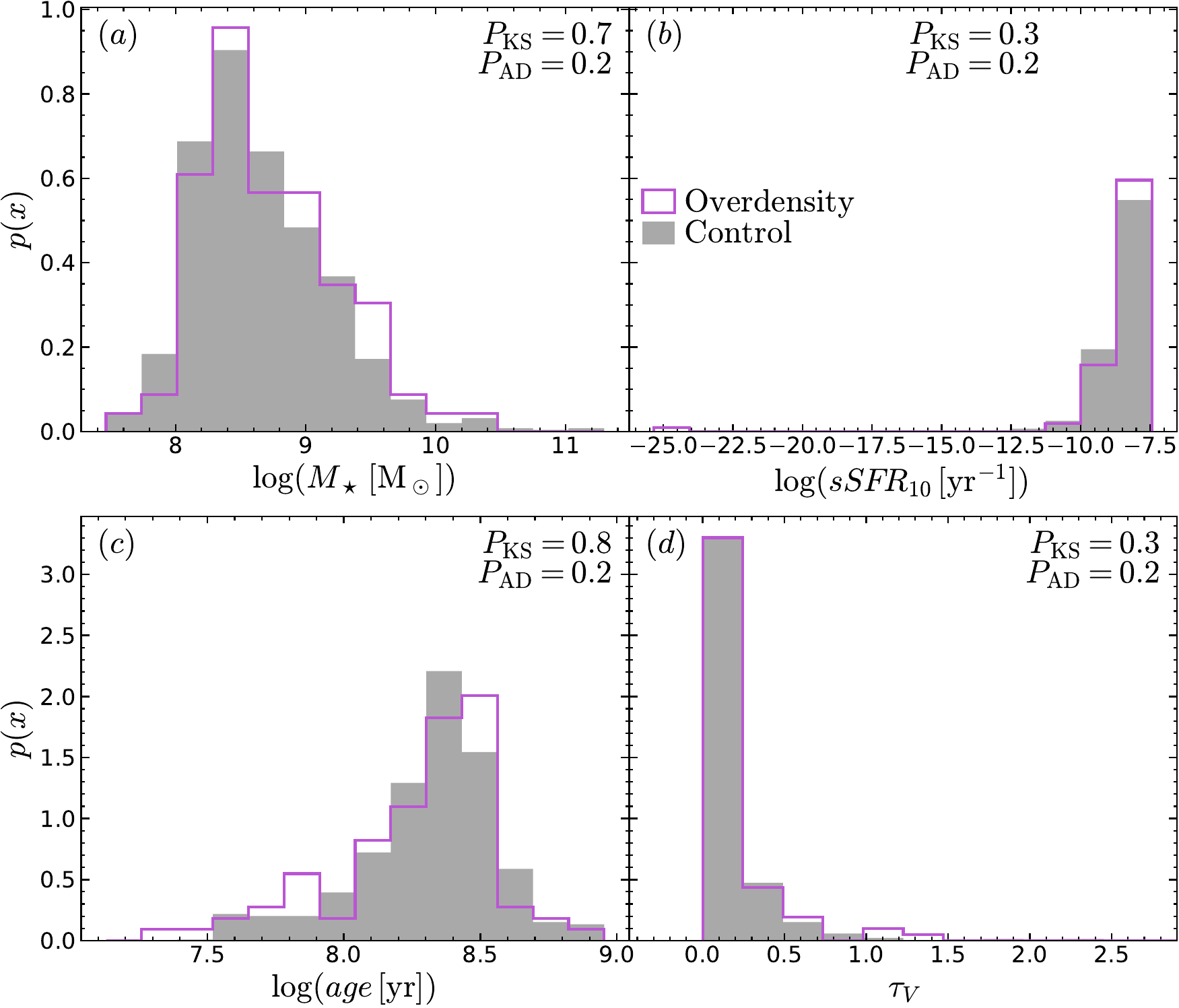}
  {\phantomsubcaption\label{f.env.phy.a}
   \phantomsubcaption\label{f.env.phy.b}
   \phantomsubcaption\label{f.env.phy.c}
   \phantomsubcaption\label{f.env.phy.d}}
  \caption{Comparing the physical properties of galaxies in the candidate overdensity (purple) to a control sample of galaxies not in any overdensity (filled grey). We report the p-value $P$ of the null hypothesis that the two samples are drawn from the same distribution (according to the KS and AD tests), but find no evidence
  for differences in the physical properties between galaxies in the overdensity and the
  control sample.
  }\label{f.env.phy}
\end{figure}

The distribution of the resulting galaxy properties is shown in Fig.~\ref{f.env.phy}. 
We find no evidence for a difference in \mstar between members and non-members of the overdensity (according to the KS and AD tests, respectively; panel~\subref{f.env.phy.a}). Similarly, we find no difference in
specific SFR (panel~\subref{f.env.phy.b}, in agreement with no
difference in F410M-F444W), and no difference in the mass-weighted age and dust attenuation properties (Fig.~\ref{f.env.phy.c}
and~\subref{f.env.phy.d}). The lack of difference in mass-weighted
age maybe due to the known difficulty of measuring accurate
mass-weighted stellar properties for young galaxies -- particularly when relying on rest-frame UV and optical wavelengths. 
Alternatively, the constraining power of our small photometric 
sample may be insufficient to overcome the constraints from the prior probabilities assumed
by \prospectorbeta.

\section{Discussion}\label{s.disc}

The discovery of a large, 25-kpc diameter emission-line nebula at $z=5.89$ underscores the
combined power of medium- and wide-band photometry to identify and characterize
extended line-emitting nebulae \citep{zhu+2024}. While there are other large emission-line regions within the 300-arcmin$^2$ area of the GOODS fields covered by JADES \citep{scholtz+2020,deugenio+2024a,zhu+2024}, this nebula is unique at such early times as $z=5.89$.
However, building a larger sample for this kind of bright and rare source requires a large-area medium-band survey and
deep follow-up with spectroscopy.

For the case of \target, while only further observations can give definitive answers, the galaxy location 
at the centre of the nebula and at a photometric redshift consistent with the nebular spectroscopic redshift leaves little 
doubt about the physical association of this system.

\subsection{Large-scale environment}\label{s.disc.ss.environment}

The galaxy and its nebula belong to a candidate large-scale galaxy overdensity, with a
significance higher than JADES-GS-OD-5.928 \citep[which has been spectroscopically confirmed;][]{helton+2024b}. \target is
located at the same redshift as the candidate overdensity and within a projected distance of 2.3~cMpc from the peak density (Section~\ref{s.environment}).
While the depth of the JADES imaging varies significantly across the survey footprint in GOODS-S, using a magnitude cut in F444W magnitude of 28.5~mag and no cut on the photometric redshift uncertainties should remove any bias due to varying survey depth. Indeed, we find that the highest-significance peak is not in the deepest region of the field \citep[which is the JADES Origins Field,][dotted grey lines in Fig.~\ref{f.overd}]{eisenstein+2023b}.
Because overdensities as large as the one considered here are far from having collapsed, we do not expect the environment-driven processes seen on Mpc scales at
$z\lesssim 2$ to be already in place \citep[e.g.,][]{ji+2018,donnari+2019,donnari+2021}. However, being part of an overdensity, the galaxy may still be subject to other effects,
such as an earlier-than-average gravitational collapse, a higher-than-average ionization fraction of the inter-galactic medium \citep{lu+2024,whitler+2024}, and higher gas metallicity.

Previous studies have found an excess of galaxies with low-EW \Halpha+\NII emission in two galaxy overdensities at $z=5.7$ \citep[][$25\pm7$~percent vs $6\pm2$~percent for non-overdensity galaxies]{morishita+2024}, but we find no difference in the 
F410M-F444W colour (tracing \Halpha EW) distribution between galaxies inside or outside the overdensity (Fig.~\ref{f.env.emp.b}).
The discrepancy could be in part due to galaxies in overdensities displaying
a broader range of EWs, due to environment-driven processes acting to both enhance and suppress star
formation -- as also discussed in \citet{morishita+2024}. Another possible explanation for the
discrepancy are different methods: \citet{morishita+2024} use deep NIRSpec and NIRCam/WFSS spectroscopy, while this work relies on photometry only.
Indeed, the observations from \citet{morishita+2024} reach $EW = 30\text{--}100$~\AA, which would
correspond to a F410M-F444W colour of $\sim 0.2$ mag. Given the survey depth of the JADES medium-depth
regions \citep{eisenstein+2023a,deugenio+2025}, this effect is too subtle to capture in individual
galaxies at the faint end of our sample (28.5~mag in F444W), and would correspond to only a 3-\textsigma colour difference at 27.5~mag.

However, \citet{morishita+2024} also find that their low-EW galaxies have evidence of Balmer breaks
(their $D_\mathrm{e}4000$ index maps very well to the definition of Balmer-break index used here).
This result is in qualitative agreement with the observations from \citet{naidu+2024}. Our
work presents statistical evidence for galaxies in the candidate overdensity having redder F200W-F277W colour than typical galaxies at the same redshift. We discussed that a difference in emission-line properties is unlikely, given the lack of difference in F410M-F444W (tracing primarily $EW(\Halpha)$), which together with the redder F200W-F277W colour would imply a systematically stronger \OIIall emission.
We therefore interpret the colour difference as evidence for galaxies in overdensities having stronger Balmer breaks (Section~\ref{f.env.emp}). This could in turn be due to older
light-weighted ages and/or higher metallicities. While higher
gas metallicities in dense environments than in the field
have been measured up to $z=0.35$
\citep{gupta+2016}, there is yet no direct evidence for metallicity enhancements at $z\sim 6$. Besides, a higher \textit{stellar}
metallicity would almost certainly imply much larger stellar masses to explain the chemical enrichment.
In contrast, older stellar population ages would naturally fit in the standard \textLambda CDM model, where over-dense regions start to collapse earlier, giving their galaxies a head start in the
SFH. Deep medium-band imaging \citep[e.g.,][]{trussler+2024} or follow-up spectroscopy \citep[e.g.,][]{looser+2024,baker+2025} are required to confirm the presence of these Balmer breaks and to
investigate their physical cause.

On the spatial scales of the CGM (tens of kpc), we may already be seeing environment-driven effects, as already witnessed at $z\sim  3.5$ \citep{alberts+2024}. The centre of the overdensity (Fig.~\ref{f.centre}) contains
six galaxies, which as a sample stand out in both F444W magnitude and F200W-F277W colour (Fig.~\ref{f.env.centre}). This suggests that large overdensities may be connected to the early formation of massive galaxies in their centres, consistent with the finding that
massive, dusty star-forming galaxies are good tracers of galaxy proto-clusters \citep{daddi+2009,riechers+2010,oteo+2018,pavesi+2018,drake+2020}, including at high redshifts \citep{smolcic+2017,lewis+2018,hashimoto+2023,arribas+2024}.
However, neither \target nor the galaxies nearest to the centre of
the overdensity are particularly dusty, so we could be witnessing
the assembly phase before the dusty starburst event.

An excess of Balmer-break galaxies in the densest environments may point to the synergy between internal feedback and environment. Star-formation or AGN-driven feedback can temporarily interrupt star formation \citep{Ceverino2018MNRAS.480.4842C,Lovell2022arXiv221107540L}, while the central galaxy could be appropriating the gas reservoir of the satellite, preventing re-accretion and rejuvenation \citep{gelli+2025}. Large-scale AGN outflows like the one seen around \target -- if they are sufficiently common -- could also impact low-mass satellites, for
instance by triggering a starburst and accelerating the depletion and/or removal of gas. \citep{croft+2006,inskip+2008,salome+2015}.

The possible presence of a satellite
(\satellite) with a strong Balmer break and a mass ratio of 1:6 seems significant (Section~\ref{s.sed.ss.satellite}). This is because galaxies with strong Balmer breaks (i.e., having had negligible star formation for hundreds of Myr) with
masses $10^9 \lesssim \mstar \lesssim 10^{10}~\Msun$ are rare at $z>3$ \citep{baker+2024,trussler+2024}. If confirmed, the coincidence would be significant,
suggesting that `environment pre-processing' may be already at work in the first billion years after the Big Bang \citep{alberts+2024}.
It is worth remarking that the best-fit \eazy photometric redshift of \satellite is $z_\mathrm{phot}=4.4$, therefore our results should be considered with caution (Section~\ref{s.sed.ss.satellite}). However, the absence of emission
lines and possible contamination of the photometry by the brighter central could be biasing this measurement. Besides, such a low-mass quiescent
galaxy, if confirmed to be at $z=4.4$, would be remarkably isolated, which would be even more extraordinary than a satellite at $z=5.89$. Either way,
spectroscopic confirmation is required.

\subsection{Main galaxy}\label{s.disc.ss.central}

The galaxy SED shows strong evidence of a break or line excess between F200W and F277W
(Fig.~\ref{f.sed.a}). If due to line emission, this would imply a rest-frame $EW(\OIIall) 
< -250~\AA$, extremely high for \OIIall \citep{blanton+lin2000,yan+2006}. If true,
such a high-magnitude EW is certainly worth studying. However, our fiducial interpretation is that the 
difference between F200W and F277W is due to a Balmer break, implying that evolved stars 
are present and dominate the mass budget.
This is supported by our SED models, which still find a Balmer break even for the highest-magnitude $EW(\OII)$ explored. 
While remarkable, a strong Balmer break at $z=5.89$
is not implausible, with the most distant such case being at $z=7.3$ \citep{weibel+2024b}, and weaker breaks
at even higher redshifts \citep{looser+2024,kuruvanthodi+2024,baker+2025}, and some $z=6$ cases even pre-dating \jwst \citep{eyles+2005,eyles+2007}.
With $\mstar \sim 10^{10}$~\Msun, \target appears to be fairly massive (for  $z\sim6$), 
only two times lower than the first massive, quiescent galaxies \citep{weibel+2024b} and comparable to the knee of the galaxy mass function at $z=6$ \citep{weaver+2023,weibel+2024a}.
We remark that \mstar is also subject to systematic uncertainties of 0.3 dex \citep[as found by e.g.,][]{muzzin+2009,conroy+2009,pacifici+2023}. These uncertainties arise due to the inability of photometry in the rest-frame UV--optical to fully break the degeneracy between age and dust, or age and metallicity \citep[e.g.,][]{trager+1998,nersesian+2024}.
As for the SFH, SED modelling suggests that the galaxy may have experienced an upturn in  
the most recent 10~Myr (Fig.~\ref{f.sed.c}). This upturn is required to explain the medium-band 
excess in F335M vs F356W, itself due to \Hbeta and \OIIIall line emission. We caution that 
by construction, our \prospector setup interprets this emission as due to star-formation 
photoionization, hence the upturn in SFR. However, AGN photoionization and shocks are equally 
plausible drivers of the observed photometric excess but are not included in our model.
For this reason, the SFR upturn in the last 10~Myr is not conclusive. To confirm if the SFR in \target is really resurrecting, we need 
3--4~\mum spectroscopy. In contrast, the downturn 10--100 Myr prior 
to observation is driven by the Balmer break and shape of the stellar continuum, so it 
should be fairly robust against AGN- and shock-driven emission. The large scatter in the
SFH during the period 10--100 Myr is due to the photometry being unable to precisely pinpoint in time
the epoch of the SFR downturn. However, the presence of a downturn seems robust
(Section~\ref{s.sed.ss.central}; Fig.~\ref{f.sed.d}). A case of a strong shock in a galaxy with a Balmer break was recently identified at $z=4.7$ by the WIDE Survey \citep{maseda+2024,deugenio+2025}. However, the similarity is not complete, because while their galaxy presents very strong \OIIall emission, its
\OIIIall emission is remarkably weak and the
SFH does not display any substantial downturn
\citep{deugenio+2025}.

\begin{figure*}
\centering
\includegraphics[width=\linewidth]{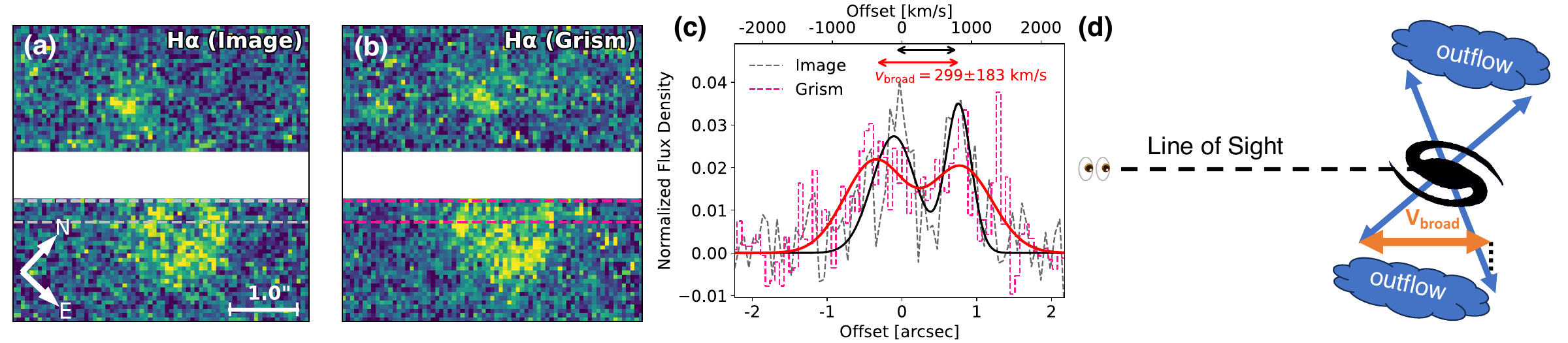}
{\phantomsubcaption\label{f.broad.a}
 \phantomsubcaption\label{f.broad.b}
 \phantomsubcaption\label{f.broad.c}
 \phantomsubcaption\label{f.broad.d}}
\caption{Evidence of kinematic broadening supports the biconical outflow interpretation for the emission-line nebula. Panel~\subref{f.broad.a}: \Halpha emission constructed from F444W--F410M image, aligned with dispersion direction.
Panel~\subref{f.broad.b}: \Halpha emission in the 2D grism spectra.
Panel~\subref{f.broad.c}: \Halpha profiles in the image (grey) and grism (pink) extracted from the apertures shown by the dashed lines in panels~\subref{f.broad.a} and~\subref{f.broad.b}.
Velocity offset in the slitless spectrum is degenerate with spatial offset.
By fitting the profiles with double Gaussian functions we obtain evidence of kinematic broadening of $\sim300~\kms$.
This is consistent with the expectation from biconical gas outflow as illustrated by the cartoon in panel~\subref{f.broad.d}.
}\label{f.broad}
\end{figure*}

\subsection{Origins of the emission-line nebula}\label{s.disc.ss.emlines}

The large physical size, velocity and high luminosity of the nebula suggest that its origin may be 
linked to AGN-driven outflows and AGN photoionization
\citep{lintott+2009}. While some extreme starbursts are also thought to drive massive, extended outflows \citep{rupke+2019}, the SFH in \target disfavours this interpretation. Still, tidal disruption or strong shocks are still possible alternatives to the
AGN scenario and will require spectroscopy \citep[e.g.,][]{deugenio+2025} and ideally integral-field spectroscopy \citep[e.g.,][]{saxena+2024,venturi+2023}
for a definitive assessment. Still, with the data in hand, several lines of evidence 
favour AGN outflows. First, the extremely compact morphology of \Hbeta+\OIII in the galaxy itself supports the
presence of a point source or extremely compact narrow-line region (Fig.~\ref{f.emlmap}).
When we consider the tentative MIR fluxes from \textit{Spitzer} and the MIRI detection at 15~\mum
(Section~\ref{s.agn}), the data are best reproduced by
a model with an obscured AGN (Fig.~\ref{f.cigale}).
The similar reduced-$\chi^2$ values between the
AGN and no-AGN models are driven by the large discrepancy around the Balmer break, likely due the SFH
parametrization we used in \cigale. Nevertheless, the
two models are formally consistent with the data, so we
consider this circumstantial evidence only.
The presence of an AGN, if confirmed, would resonate with recent outflows, though
admittedly mergers and tidal disruption could also trigger an AGN 
\citep{perna+2023b,ellison+2024}. The presence of two clouds symmetrically located on 
either side of the galaxy also favours an AGN origin, as do their convex shapes, 
reminiscent of bow shocks and bubbles \citetext{\citealp{nelson+2019}; see e.g., \citealp{cresci+2023,veilleux+2023,venturi+2023} for observational evidence}.
From the grism spectrum, we find some indications that the \Halpha emission from the East cloud is broader than in the NIRCam image (cf. pink vs black lines in Fig.~\ref{f.broad.c}). This additional broadening corresponds to a velocity dispersion of $\sim300~\kms$, consistent with the expected kinematic broadening from conical outflow geometry (Fig.~\ref{f.broad}).
However, this will need to be further confirmed with NIRSpec/IFS data at higher spectral resolution and sensitivity.
A final argument in favour of an AGN is the high
\Halpha luminosity; in the merger/tidal disruption scenario, the \Halpha luminosity would 
imply a very high SFR of 80~\Msun~\peryr in the nebula, which should be associated with
easily detectable stellar continuum light, yet very little continuum is seen. Extreme emission-line galaxies
are generally found to be compact \citep{whiters+2023,boyett+2024}, unlike our case.
The SFR value is very large for a tidally disrupted satellite; for comparison, the extended emission-line
region in the Tusitala group, around the much more massive galaxy COS20115
\citep[`Jekyll', $\mstar = (0.9\text{--}1.8)\times10^{11}~\Msun$;][]{glazebrook+2017,
schreiber+2018,perez-gonzalez+2024} has a total equivalent SFR that is three times lower,
raising the question of whether \target, which is ten times less massive than Jekyll,
could disrupt three times more gas. Moreover, the high line-to-continuum ratio in the clouds (which SED modelling struggles to reproduce, Section~\ref{s.sed.ss.clouds}) imply extremely young mass-weighted stellar ages (of order $\sim 1$~Myr). These young ages are implausible over such an extended region, because the speed of the signal needed to synchronize star formation over 10~kpc would be of order $10^4~\kms$. This is much larger than the outflow velocities typically measured on comparable scales. 

While the tidal disruption explanation seems unlikely, our favoured AGN-outflows scenario
is not without problems. Powerful outflows are generally associated with intense accretion
activity, but \target appears to be a fairly faint, spatially resolved target, unlike bright
and point-like quasars.
A similar but less extended and luminous case has been recently
found near the faint QSO HSC J2239+0207 \citep{lyu+2024}. The comparison
with \citet{lyu+2024} makes our nebula even more puzzling, since
their nebula is less extended and luminous, while the central galaxy is at least a QSO, unlike \target.
The MIR constraints place an upper limit on the AGN
luminosity of $L_\mathrm{bol} \lesssim 10^{45.8} \, \mathrm{erg\;s^{-1}}$, significantly lower than bright
quasars.
Still, observations from the local Universe show that AGN can fade by orders of magnitude over timescales of a few Myr \citep{finlez+2022}, so even though \target is not a bright QSO, we could be observing a relic nebula \citep{lintott+2009}.
Future MIRI observations are necessary to place more
stringent constraints on the obscured AGN.

\subsection{A link with quenching?}\label{s.disc.ss.quenching}

Intriguingly, SED modelling suggests that the galaxy may have experienced
a downturn in recent star formation (Fig.~\ref{f.sed.c}). If confirmed, this scenario may
be linked to the presence of the large nebula. In the absence of strong ionizing emission
(as suggested by the faint nature of \target), the lifetime of collisionally excited 
emission should be fairly short, of order $10^4$~yr. In addition, the velocity offset 
between the East and West clouds is $\Delta\,v \approx 800~\kms$ (Section~\ref{s.redshift}). Interpreting the clouds as originating from the
galaxy, and using the projected separation of 6~kpc between \target and the highest-brightness cloud, we can derive an outflow age of at least $\approx 15$~Myr, clearly longer than the cooling time, and thus requiring external photoionization (for example, from a faded AGN in \target), or a local source of power like fast shocks.
The age of the putative outflow matches well the recent upturn in the SFH (Fig.~\ref{f.sed.d}).

Could massive outflow events be linked to galaxies quenching? Most definitions of quiescence require no star formation for hundreds of Myr; assuming the visibility time for the cloud to be 15~Myr, galaxies like \target could be the progenitors of 10$\times$ 
larger volume densities of quiescent galaxies,
meaning that extended, luminous nebulae
may be a common but short-lived evolutionary phase around massive galaxies at these high redshifts. The presence of possible signatures of
a recent merger (Section~\ref{s.morph}) resonate with
the merger-then-quenching scenario \citep{hopkins+2008}, where a major merger
increases both the SFR and the accretion rate of the
supermassive black hole until a `blowout' phase that precedes quenching. If we are truly witnessing this short-lived blowout phase in \target, this would have implications for
our understanding of how SMBH feedback affects the host galaxy and causes quenching.
So far there are only few observations of the circum-galactic medium around massive, quiescent
galaxies at $z>3$; however, where data is sufficiently deep, all these systems appear to have
some more or less extended emission-line regions \citetext{\citealp{deugenio+2024a},
\citealp{perez-gonzalez+2024} using NIRSpec/IFS; R. Pascalau (in~prep.) using NIRCam}. This may
point to a connection between these extended emission-line regions and neutral-gas outflows, which
have been confirmed in several quiescent galaxies around and before cosmic noon \citep{davies+2024},
including the galaxy from \citet{deugenio+2024a}, where NIRSpec/IFS confirms the simultaneous presence
of both extended ionized-gas emission and neutral outflows with high mass-outflow rate \citep{scholtz+2024}.

Unfortunately, distinguishing between the scenarios presented earlier is impossible with
photometry alone, but with JWST integral field spectroscopy capable of revealing the chemical and kinematic
properties of the nebula, the key to unlocking many of the open questions surrounding \target
lies within reach.

\section{Summary and conclusions}\label{s.conc}

In this work, we report the discovery of a large and luminous line-emitting nebula around a massive galaxy at $z=5.89$ (Fig.~\ref{f.img}). The galaxy and nebula have consistent photometric redshifts, driven by the \Lyalpha break and photometric excess driven by strong emission lines. The nebula is confirmed spectroscopically using NIRCam/WFSS from SAPPHIRES (Fig.~\ref{f.grism}), which detects \Halpha at a mean redshift $z_\mathrm{spec}=5.89\pm0.01$.
\begin{itemize}
  \item SED modelling with \prospector (Fig.~\ref{f.sed.b}) shows that the central galaxy \target has a stellar mass
  $\log(\mstar/\Msun) = 10.1^{+0.1}_{-0.2}$, near the knee of the
  mass function. In addition, the SFH shows a drop in star formation (10--100~Myr before observation, causing the observed Balmer break) followed by a recent upturn (powering the emission lines;
  Fig~\ref{f.sed.d}).
  \item The emission-line regions cannot be modelled satisfactorily with our \prospector setup, implying these are gas clouds and
  not galaxies (Fig.~\ref{f.sed.e} and~\subref{f.sed.j}).
  \item The galaxy has a possible nearby satellite \satellite (Figs.~\ref{f.img.a} and~\ref{f.sed.g}).
  If confirmed, this would have a strong Balmer break (Fig.~\ref{f.sed.i}). Its coincidence with \target would imply short-range interactions can enable or even cause quenching at $z\gtrsim5.89$.
  \item SED modelling with \cigale does not require an AGN, but places a limit on the AGN luminosity of
  $\log (L_\mathrm{bol} \, [\mathrm{erg\;s^{-1}}]) < 45.8\pm0.6$~dex.
  \item The size, shape, kinematics, and high luminosity of the nebula ($L_{\OIIIL} = 4.6\times10^{43}$~erg~s$^{-1}$, Section~\ref{s.eml})
  suggest a link with AGNs.
  \item \target is found $\sim2$~cMpc from the centre of a candidate
  galaxy overdensity (Fig.~\ref{f.overd}). This coincidence suggests
  a link between the massive nature of the galaxy and its membership into the overdensity (Section~\ref{s.environment.ss.overd}).
  \item When compared to a control sample of coeval galaxies, the members of the overdensity have redder F200W-F277W colour, suggesting a stronger Balmer break strength and pointing
  to an earlier formation of galaxies in the overdensity (Section~\ref{s.environment.ss.effects}, Figs.~\ref{f.env.emp}).
  \item We speculate about a possible link between the presence of such a remarkable nebula and the downturn in SFR, possible
  evidence of a short-lived `blowout' phase before the onset of quiescence.
\end{itemize}
Future observations with \jwst are required to fully understand the nature of this source, and its significance in the broader context of galaxy evolution.




\section*{Acknowledgements}
FDE, RM, IJ, GCJ, DP and JS acknowledge support by the Science and Technology Facilities Council (STFC), by the ERC through Advanced Grant 695671 ``QUENCH'', and by the
UKRI Frontier Research grant RISEandFALL. RM also acknowledges funding from a research professorship from the Royal Society.
SA and MP acknowledge grant PID2021-127718NB-I00 funded by the Spanish Ministry of Science and Innovation/State Agency of Research (MICIN/AEI/ 10.13039/501100011033).
MP also acknowledges support from the Programa Atracci\'on de Talento de la Comunidad de Madrid via grant 2018-T2/TIC-11715.
SC and GV acknowledge support by European Union's HE ERC Starting Grant No. 101040227 - WINGS.
ECL acknowledges support of an STFC Webb Fellowship (ST/W001438/1).
ALD thanks the University of Cambridge Harding Distinguished Postgraduate Scholars Programme and Technology Facilities Council (STFC) Center for Doctoral Training (CDT) in Data intensive science at the University of Cambridge (STFC grant number 2742605) for a PhD studentship.
EE, DJE, BER and YZ acknowledge the JWST/NIRCam contract to the University of Arizona NAS5-02015.
DJE is supported as a Simons Investigator.
YF is supported by JSPS KAKENHI Grant Numbers JP22K21349 and JP23K13149.
BER acknowledges support from JWST Program 3215.
AJB acknowledges funding from the ``FirstGalaxies'' Advanced Grant from the European Research Council (ERC) under the European Union's Horizon 2020 research and innovation program (Grant agreement No. 789056).
IJ and DP also acknowledge support by the Huo Family Foundation through a P.C. Ho PhD Studentship.
PGP-G acknowledges support from grant PID2022-139567NB-I00 funded by Spanish Ministerio de Ciencia e Innovaci\'on MCIN/AEI/10.13039/501100011033, FEDER, UE.
ST acknowledges support by the Royal Society Research Grant G125142.
The research of CCW is supported by NOIRLab, which is managed by the Association of Universities for Research in Astronomy (AURA) under a cooperative agreement with the National Science Foundation.
H\"U acknowledges funding by the European Union (ERC APEX, 101164796). Views and opinions expressed are however those of the authors only and do not necessarily reflect those of the European Union or the European Research Council Executive Agency. Neither the European Union nor the granting authority can be held responsible for them.
JAAT acknowledges support from the Simons Foundation and JWST program 3215. Support for program 3215 was provided by NASA through a grant from the Space Telescope Science Institute, which is operated by the Association of Universities for Research in Astronomy, Inc., under NASA contract NAS 5-03127.

This work made use of the High Performance Computing (HPC) resources at the University of Arizona, which are funded by the Office of Research Discovery and Innovation (ORDI), Chief Information Officer (CIO) and University Information Technology Services (UITS). This work made extensive use of the freely available \href{http://www.debian.org}{Debian GNU/Linux} operative system.
We used the \href{http://www.python.org}{Python} programming language \citep{vanrossum1995}, maintained and distributed by the Python Software Foundation. We made direct use of Python packages
{\sc \href{https://pypi.org/project/astropy/}{astropy}} \citep{astropyco+2013},
{\sc \href{https://pypi.org/project/corner/}{corner}} \citep{foreman-mackey2016},
{\sc \href{https://pypi.org/project/emcee/}{emcee}} \citep{foreman-mackey+2013},
{\sc \href{https://pypi.org/project/jwst/}{jwst}} \citep{alvesdeoliveira+2018},
{\sc \href{https://pypi.org/project/matplotlib/}{matplotlib}} \citep{hunter2007},
{\sc \href{https://pypi.org/project/numpy/}{numpy}} \citep{harris+2020},
{\sc \href{https://pypi.org/project/ppxf/}{ppxf}} \citep{cappellari+emsellem2004, cappellari2017, cappellari2022},
{\sc \href{https://pypi.org/project/astro-prospector/}{prospector}} \citep{johnson+2021} \href{https://github.com/bd-j/prospector}{v2.0},
{\sc \href{https://pypi.org/project/python-fsps/}{python-fsps}} \citep{johnson_pyfsps_2023} and 
{\sc \href{https://pypi.org/project/scipy/}{scipy}} \citep{jones+2001}.
We also used the softwares {\sc \href{https://github.com/cconroy20/fsps}{fsps}} \citep{conroy+2009,conroy_gunn_2010}, {\sc \href{https://www.star.bris.ac.uk/~mbt/topcat/}{topcat}}, \citep{taylor2005}, {\sc \href{https://github.com/ryanhausen/fitsmap}{fitsmap}} and {\sc \href{https://sites.google.com/cfa.harvard.edu/saoimageds9}{ds9}} \citep{joye+mandel2003}.

\section*{Data Availability}


This work is based on observations made with the NASA/ESA/CSA James Webb Space Telescope. Raw data were obtained from the \href{https://mast.stsci.edu/portal/Mashup/Clients/Mast/Portal.html}{Mikulski Archive for Space Telescopes} at the Space Telescope Science Institute, which is operated by the Association of Universities for Research in Astronomy, Inc., under NASA contract NAS 5-03127 for JWST. These observations are associated with programmes PID~1180, 1210, 1286, 1287, 1895, 1963, 3215, 4540, and 6434. A new public release of the reduced images and
photometric catalogues from PIDs 1180, 1210, 1286, 1287 and 3215 will be presented in JADES Collaboration (in~prep.).



\bibliographystyle{config/mnras}
\bibliography{astro} 

\appendix

\section{Line fluxes from multi-band photometry}\label{s.linephot}

Here we derive the approximate expressions for the 
equivalent width $EW$ and line flux  $F_{\rm line}$ of an 
emission line at observed wavelength $\lambda_0$, for the 
two cases where the line is observed in two overlapping 
filters (Eqs.~\ref{eq.ewo3} and~\ref{eq.o3sb}) or in only
one of two overlapping filters (Eqs.~\ref{eq.ewha} 
and~\ref{eq.hasb}). We denote the two filters as `a' and 
`b', their pivot wavelengths as $\lambda_\mathrm{a}$ and
$\lambda_\mathrm{b}$, and their transmission curves as
$t_\mathrm{a}$ and $t_\mathrm{b}$. We further define the
rectangular filter width as
\begin{equation}\label{eq.rekt}
  \rekt{\mathrm{a}} \equiv \dfrac{1}{T_\mathrm{a}} \int t_\mathrm{a}(\lambda) d\lambda,
\end{equation}
where $T_\mathrm{a}$ is the maximum of $t_\mathrm{a}$.
We also assume that $t_\mathrm{a}(\lambda_0)$, the
filter transmission at the wavelength of the line, is
always approximately equal to $T_\mathrm{a}$, which
unless applies to our observations, but is not necessarily
true in general. The spectral flux density can be written
as
\begin{equation}
  f_\lambda(\lambda) = s(\lambda) + l(\lambda)
\end{equation}
where $s$ and $l$ are the flux densities of the continuum 
and of the emission line, the latter being negligible a 
few FWHM away from $\lambda_0$. The weighted flux density through filter a is
\begin{equation}\label{eq.fluxd}
  F_{\nu,\mathrm{a}} \equiv \dfrac{\lambda_\mathrm{a}^2}{\rm c} \dfrac{\int t_\mathrm{a}(\lambda) \left( s(\lambda) + l(\lambda) \right) d\lambda}{\int t_\mathrm{a}(\lambda) d\lambda}.
\end{equation}

From the definition of EW, we have
\begin{equation}
  EW \equiv - \dfrac{1}{1+z} \displaystyle \int \dfrac{l(\lambda)}{s(\lambda)}d \lambda,
\end{equation}
which we can approximate as
\begin{equation}
  EW \approx - \dfrac{1}{(1+z) t(\lambda_0)} \displaystyle \int \dfrac{t(\lambda) l(\lambda)}{s(\lambda)}d \lambda,
\end{equation}
because the \jwst filters are reasonably well approximated 
by a top-hat function, and the factor $t(\lambda_0)$ 
ensures that the transmission value at $\lambda_0$ does
not change the integral. We now substitute $l(\lambda) = s(\lambda) + l(\lambda) - s(\lambda)$, obtaining
\begin{equation}
  EW \approx - \dfrac{1}{(1+z) t(\lambda_0)} \displaystyle \int \dfrac{t(\lambda) \left( l(\lambda) + s(\lambda) \right)}{s(\lambda)}d \lambda + \dfrac{ \int t(\lambda) d\lambda}{(1+z) t(\lambda_0)}.
\end{equation}
Assuming the continuum to be constant $s(\lambda) = s_0$,
we can write 
\begin{equation}
  EW \approx - \dfrac{1}{(1+z) t(\lambda_0) s_0} \int t(\lambda) \left( l(\lambda) + s(\lambda) \right) d \lambda + \dfrac{\int t(\lambda) d \lambda}{(1+z) t(\lambda_0)},
\end{equation}
and recalling the definitions of $\rekt{a}$ and $F_{\nu,\mathrm{a}}$, Eqs.~\ref{eq.rekt} and~\ref{eq.fluxd}, we obtain
\begin{equation}
  EW \approx - \dfrac{\mathrm{c} \rekt{\mathrm{a}} T_\mathrm{a}}{(1+z) t(\lambda_0) \lambda_\mathrm{a}^2 s_0} F_{\nu,\mathrm{a}} + \dfrac{\rekt{\mathrm{a}} T_\mathrm{a}}{(1+z) t(\lambda_0)}.
\end{equation}
We rearrange for convenience, and substitute $T_\mathrm{a} \approx t(\lambda_0)$, which is appropriate if the line is
well centred in the filter a, and follows from the 
assumption that the NIRCam filters are approximately shaped
as top-hat functions. We thus obtain
\begin{equation}\label{eq.ewstart}
  \dfrac{(1+z)}{\rekt{\mathrm{a}}}EW - 1 \approx - \dfrac{\mathrm{c}}{\lambda_\mathrm{a}^2} \dfrac{F_{\nu,\mathrm{a}}}{s_0}.
\end{equation}
From here, we can obtain Eq.~\ref{eq.ewo3} by dividing Eq.~\ref{eq.ewstart} by the corresponding expression for
filter b, assuming a and b to be the F335M and F356W filters, and solving for $EW$. In this case, since
we assumed $s(\lambda)=s_0$ to be constant, we also have
$\delta_\mathrm{cont} = 1$.
To obtain Eq.~\ref{eq.ewha}, where F410M does not contain
the emission line, we can set filter a to be F444W, and
substitute
\begin{equation}
s_0 = \dfrac{\mathrm{c}}{\lambda_\mathrm{F410M}^2} F_{\nu,\mathrm{F410M}}.
\end{equation}
The correction terms for the case of continuum that is constant in $f_\nu$ can be obtained by replacing $s(\lambda) = f_0 \mathrm{c} / \lambda^2$. In this case,
instead of Eq.~\ref{eq.ewstart}, we obtain
\begin{equation}
  \dfrac{(1+z)}{\rekt{\mathrm{a}}}EW - 1 \approx - \dfrac{F_{\nu,\mathrm{a}}}{F_{\nu,\mathrm{a},\mathrm{cont}}},
\end{equation}
from which we can derive Eqs.~\ref{eq.ewo3} and~\ref{eq.ewha} in a similar fashion as done for the
case when the continuum is constant in $f_\lambda$.

For the emission-line fluxes, Eqs.~\ref{eq.o3sb} and~\ref{eq.hasb}, we start again from Eq.~\ref{eq.fluxd}.
Assuming the continuum $s(\lambda) = s_0$ to be constant, the first 
addend can be written as
\begin{equation}
  \dfrac{\lambda_\mathrm{a}^2}{\rm c} \dfrac{\int t_\mathrm{a}(\lambda) s(\lambda) d\lambda}{\int t_\mathrm{a}(\lambda) d\lambda} = \dfrac{\lambda_\mathrm{a}^2}{\rm c} s_0.
\end{equation}
This term is instead equal to simply $s_0$ for the case
where the continuum $s$ is constant in $f_\nu$, since
there would be an additional factor $\mathrm{c}/\lambda_\mathrm{a}^2$ that simplifies the original expression. As for the second addend in Equation~\ref{eq.fluxd}, we have
\begin{equation}
  \dfrac{\lambda_\mathrm{a}^2}{\rm c} \dfrac{\int t_\mathrm{a}(\lambda) l(\lambda) d\lambda}{\int t_\mathrm{a}(\lambda) d\lambda} \approx \alpha_\mathrm{a} \dfrac{\lambda_\mathrm{a}^2}{\rm c} \dfrac{F_{\rm line} t_\mathrm{a}(\lambda_0)}{\rekt{\mathrm{a}} T_\mathrm{a}} \approx 
  \alpha_\mathrm{a} \dfrac{\lambda_\mathrm{a}^2}{\rm c} \dfrac{F_{\rm line}}{\rekt{\mathrm{a}}}.
\end{equation}
In these expressions, $\alpha_\mathrm{a}$ is a space-saving notation,
taken to be $\alpha_\mathrm{a} = 1$ if the emission line is observed
in filter `a', and 0 otherwise. 

Summarizing, and considering both filters $a$ and $b$, we have
\begin{equation}\label{eq.sb}
\begin{split}
  F_{\nu,\mathrm{a}} & = \alpha_\mathrm{a} \dfrac{\lambda_\mathrm{a}^2}{\rm c \; \rekt{\mathrm{a}}} F_{\rm line} + \dfrac{\lambda_\mathrm{a}^2}{\rm c} s_0\\
  F_{\nu,\mathrm{b}} & = \alpha_\mathrm{b} \dfrac{\lambda_\mathrm{b}^2}{\rm c \; \rekt{\mathrm{b}}} F_{\rm line} + \dfrac{\lambda_\mathrm{b}^2}{\rm c} s_0.
\end{split}
\end{equation}
Eqs.~\ref{eq.ewo3} and~\ref{eq.hasb} can be
obtained from Eq.~\ref{eq.sb} by considering
\begin{equation}
  F_{\nu,\mathrm{a}} - \dfrac{\lambda_\mathrm{a}^2}{\lambda_\mathrm{b}^2} F_{\nu,\mathrm{b}} = \dfrac{\lambda_\mathrm{a}^2}{\rm c} F_{\rm line} \left(\dfrac{\alpha_\mathrm{a}}{\rekt{\mathrm{a}}} - \dfrac{\alpha_\mathrm{b}}{\rekt{\mathrm{b}}}\right)
\end{equation}
and solving for $F_{\rm line}$. For Eq.~\ref{eq.o3sb}, the
filters a and b are F335M and F356W, so $\alpha_\mathrm{a}=1=\alpha_\mathrm{b}$, while for
Eq.~\ref{eq.hasb}, a and b are F444W and F410M,  so $\alpha_\mathrm{a}=1$ and $\alpha_\mathrm{b}=0$.

\section{Centre of the overdensity}\label{s.centre}

\begin{figure}
  \includegraphics[width=\columnwidth]{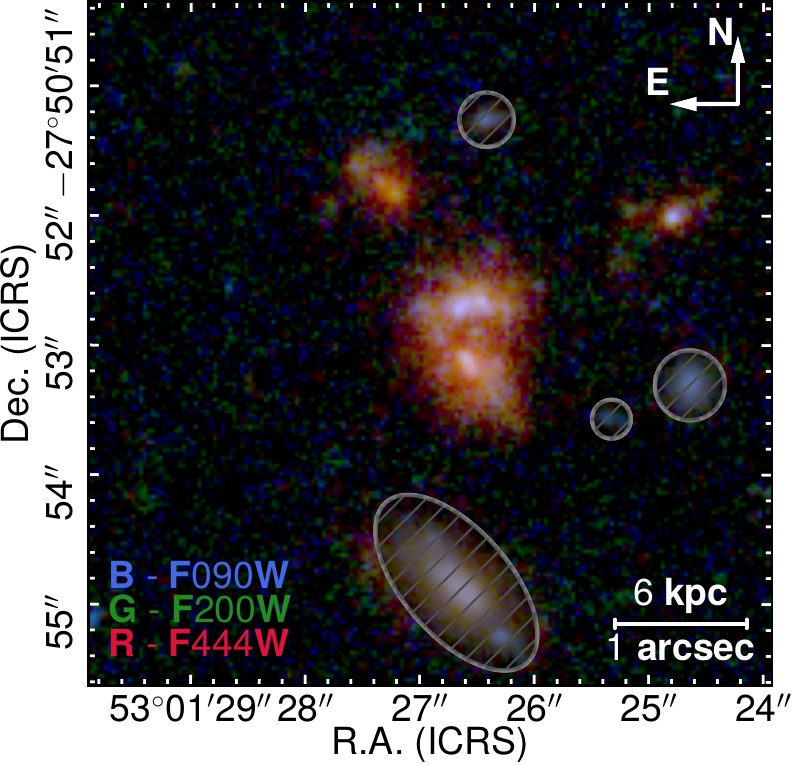}
  \caption{NIRCam false-colour image of the 5-arcsec square near the centre of the candidate overdensity (Fig.~\ref{f.overd}). We find
  five galaxies with $z_\mathrm{phot}$ consistent with the redshift
  of the overdensity. All these galaxies have positive F200W-F277W
  (Fig.~\ref{f.env.centre.c}).}\label{f.centre}
\end{figure}

\begin{figure}
  \includegraphics[width=\columnwidth]{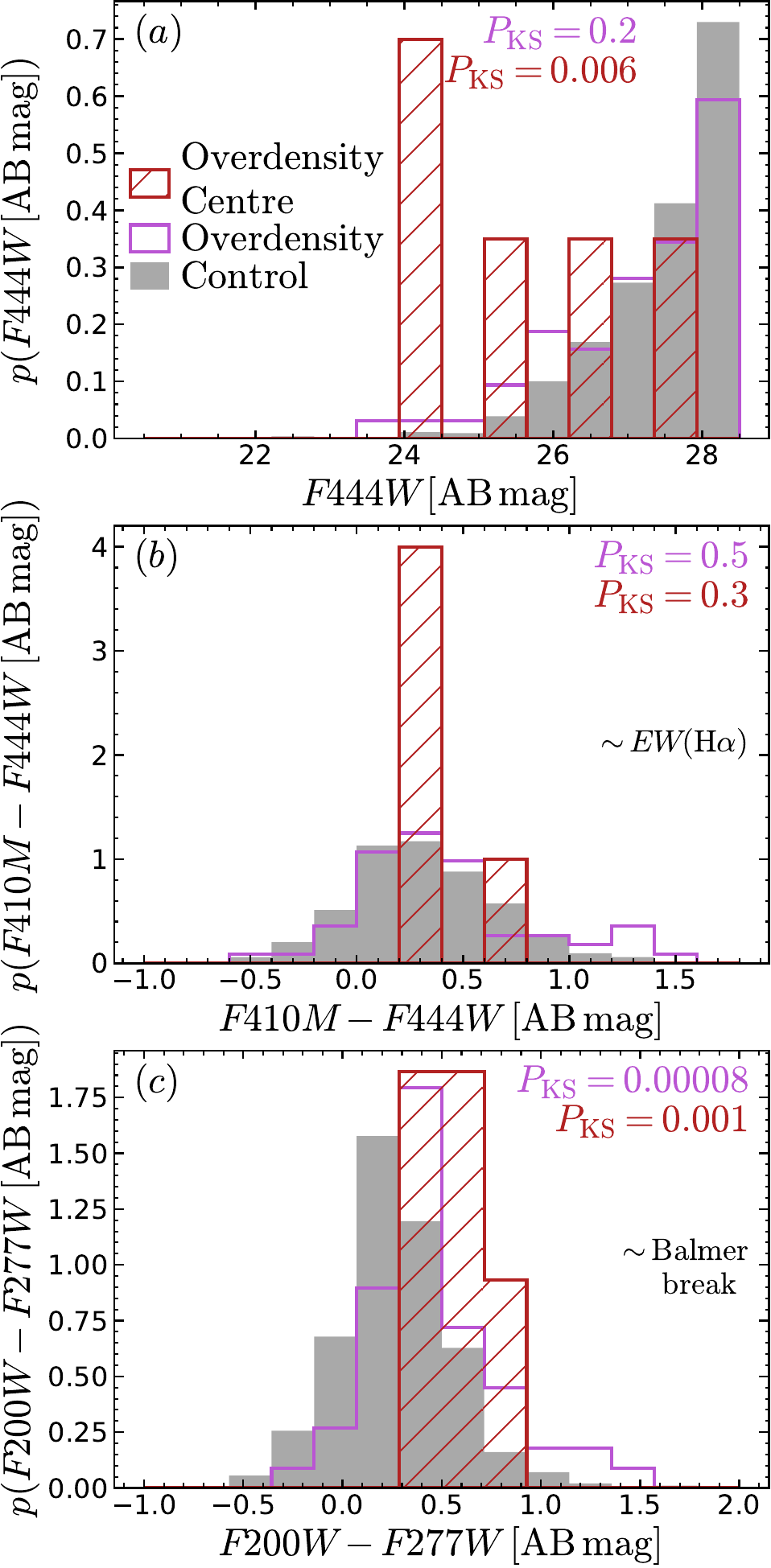}
  {\phantomsubcaption\label{f.env.centre.a}
   \phantomsubcaption\label{f.env.centre.b}
   \phantomsubcaption\label{f.env.centre.c}}
  \caption{Same as Fig.~\ref{f.env.emp}, but also showing the five galaxies nearest to the centre of the overdensity (red hatched histogram). Despite the much smaller sample size, we still find a statistically significant difference in the F200W-F277W colour
  (panel~\subref{f.env.centre.c}), pointing to small-scale
  environment having a distinct effect on the SFH of galaxies.
  }\label{f.env.centre}
\end{figure}

The sources nearest to the centre of the candidate overdensity form
a close group of at least five distinct galaxies, all within 1.5~arcsec from the source JADES-GS-429475.
Based on their $z_\mathrm{phot}$ and location, all these sources belong to the candidate overdensity.
JADES-GS-429475 is located 17.5~arcsec (103~pkpc) from the
centre of the candidate overdensity. The presence of several galaxies at the same redshift and in such a small area of the sky
(a square of less than 3-arcsec side, Fig.~\ref{f.centre}) suggests that the centre of large-scale overdensities may be connected to
the formation of massive, early galaxies.
We find some evidence for these galaxies being brighter than the
typical galaxy at $z\sim 6$ (Fig.~\ref{f.env.centre.a}; 2.5~\textsigma). The fact that all five galaxies have positive
F200W-F277W colour seems significant, given a sample size of
only five (3~\textsigma, Fig.~\ref{f.env.centre.c}).
This excess of Balmer-break galaxies may point to older stellar populations, likely due to an earlier start of the SFH.

\bsp	
\label{lastpage}
\end{document}